\documentclass{aa}

\usepackage{lscape}
\usepackage{graphicx}
\usepackage[varg]{txfonts}
\usepackage[colorlinks=true, allcolors=blue]{hyperref}

\usepackage{amsmath}
\usepackage{amssymb}
\usepackage{textcomp}
\usepackage{gensymb}
\usepackage{mathtools}
\usepackage{subcaption}
\usepackage{multirow}
\usepackage{colortbl}
\usepackage{xcolor}
\usepackage{arydshln}
\usepackage[switch]{lineno}

\newcommand{\dok}{\citetalias{2021A&A...651A..86D}}
\defcitealias{2021A&A...651A..86D}{D21}

\newcommand{\loren}{\citetalias{2017A&A...605A..58A}}
\defcitealias{2017A&A...605A..58A}{A17}

\makeatletter
\renewcommand*\aa@pageof{, page \thepage{} of \pageref*{LastPage}}
\makeatother

\begin{document}

\title{A global view on star formation: The GLOSTAR Galactic plane survey}
\subtitle{VII. Supernova remnants in the Galactic longitude range $28\degree<l<36\degree$}
\titlerunning{GLOSTAR: SNRs II}

\author{R.\,Dokara \inst{1}\fnmsep\thanks{Member of the International Max Planck Research School (IMPRS) for Astronomy and Astrophysics at the Universities of Bonn and Cologne}
        \and
        Y.\,Gong \inst{1}
        \and
        W.\,Reich \inst{1}
        \and
        M.\,R.\,Rugel \inst{1,2,3}\fnmsep\thanks{Jansky Fellow of the National Radio Astronomy Observatory}
        \and
        A.\,Brunthaler \inst{1}
        \and
        K.\,M.\,Menten \inst{1}
        \and
        W.\,D.\,Cotton \inst{4,5}
        \and
        S.\,A.\,Dzib \inst{6,1}
        \and
        S.\,Khan \inst{1,\star}
        \and
        S.\,-N.\,X.\,Medina \inst{1,7}
        \and
        H.\,Nguyen \inst{1,\star}
        \and
        G.\,N.\,Ortiz-Le{\'o}n \inst{8,1}
        \and
        J.\,S.\,Urquhart \inst{9}
        \and
        F. Wyrowski \inst{1}
        \and
        A.\,Y.\,Yang \inst{1}
        \and
        L.\,D.\,Anderson \inst{10}
        \and
        H.\,Beuther \inst{11}
        \and
        T.\,Csengeri \inst{12}
        \and
        P.\,M\"uller \inst{1}
        \and
        J.\,Ott \inst{3}
        \and
        J.\,D.\,Pandian \inst{13}
        \and
        N.\,Roy \inst{14}
        }
\authorrunning{R. Dokara et al.}

\institute{Max Planck Institute for Radioastronomy (MPIfR), Auf dem H\"ugel 69, 53121 Bonn, Germany\\
           \email{rdokara@mpifr-bonn.mpg.de}
           \and
           Center for Astrophysics, Harvard \& Smithsonian, 60 Garden St., Cambridge, MA 02138, USA
           \and
           National Radio Astronomy Observatory, 1003 Lopezville Rd, Socorro, NM 87801, USA
           \and
           National Radio Astronomy Observatory, 520 Edgemont Road, Charlottesville, VA, 22903, USA
           \and
           South African Radio Astronomy Observatory, 2 Fir St, Black River Park, Observatory, 7925, South Africa
           \and
           IRAM, 300 rue de la piscine, 38406 Saint Martin d'H\`eres, France
           \and
           German Aerospace Center, Scientific Information, 51147 Cologne, Germany
           \and
           Instituto de Astronom\'ia, Universidad Nacional Aut\'onoma de M\'exico (UNAM), Apdo Postal 70-264, Ciudad de M\'exico, M\'exico
           \and
           Centre for Astrophysics and Planetary Science, University of Kent, Ingram Building, Canterbury, Kent CT2 7NH, UK
           \and
           Department of Physics and Astronomy, West Virginia University, Morgantown, WV, 26506, USA
           \and
           Max Planck Institute for Astronomy, K\"{o}nigstuhl 17, D-69117 Heidelberg, Germany
           \and
           Laboratoire d'astrophysique de Bordeaux, Univ. Bordeaux, CNRS, B18N, all\'ee Geoffroy Saint-Hilaire, 33615 Pessac, France
           \and
           Department of Earth \& Space Sciences, Indian Institute of Space Science and Technology, Trivandrum 695547, India
           \and
           Department of Physics, Indian Institute of Science, Bengaluru 560012, India
           }

\date{Received ...; accepted ...}



\abstract
{ While over 1000 supernova remnants (SNRs) are estimated to exist in the Milky Way, only less than 400 have been found to date.  In the context of this apparent deficiency, more than 150 SNR candidates were recently identified in the D-configuration Very Large Array (VLA-D) continuum images of the 4--8~GHz global view on star formation (GLOSTAR) survey, in the Galactic longitude range $-2\degree<l<60\degree$. }
{ We attempt to find evidence of nonthermal synchrotron emission from 35 SNR candidates in the region of Galactic longitude range $28\degree<l<36\degree$, and also to study the radio continuum emission from the previously confirmed SNRs in this region. }
{ Using the short-spacing corrected GLOSTAR VLA-D+Effelsberg images, we measure the ${\sim}6$~GHz total and linearly polarized flux densities of the SNR candidates and the SNRs that were previously confirmed.  We also attempt to determine the spectral indices by measuring flux densities from complementary Galactic plane surveys and from the temperature-temperature plots of the GLOSTAR-Effelsberg images. }
{ We provide evidence of nonthermal emission from four candidates that have spectral indices and polarization consistent with a SNR origin, and, considering their morphology, we are confident that three of these (G28.36+0.21, G28.78-0.44, and G29.38+0.10) are indeed SNRs.  However, about $25\%$ of the candidates (8 out of 35) have spectral index measurements that indicate thermal emission, and the rest of them are too faint to have a good constraint on the spectral index yet.  }
{ Additional observations at longer wavelengths and higher sensitivities will shed more light on the nature of these candidates.  A simple Monte-Carlo simulation reiterates the view that future studies must persist with the current strategy of searching for SNRs with small angular size to solve the problem of the Milky Way's missing SNRs. }


\maketitle

\section{Introduction}

The structure formed from the expelled material and the shock-wave of a supernova explosion interacting with the surrounding interstellar medium (ISM) is known as a supernova remnant (SNR). The interactions of expanding SNRs and the ISM are important feedback mechanisms that may trigger star formation or, on the contrary, disperse gas and thus 
suppress the star formation rate in a galaxy.  Gas can be blown out of the Galactic plane, and  turbulent pressure is produced and maintained on both small (molecular cloud) and large (galaxy-wide) scales \citep[e.g. ][]{2000MNRAS.317..697E,2011ApJ...731...41O,2015A&ARv..23....3D,2020A&A...641A..70B}.  In order to fully understand and quantify the impact SNRs have on the dynamics of star formation in the Milky Way from an observational point of view, having a complete catalog of Galactic SNRs is highly desirable.

\begin{figure*}[ht]
    \centering
    \includegraphics[width=\textwidth]{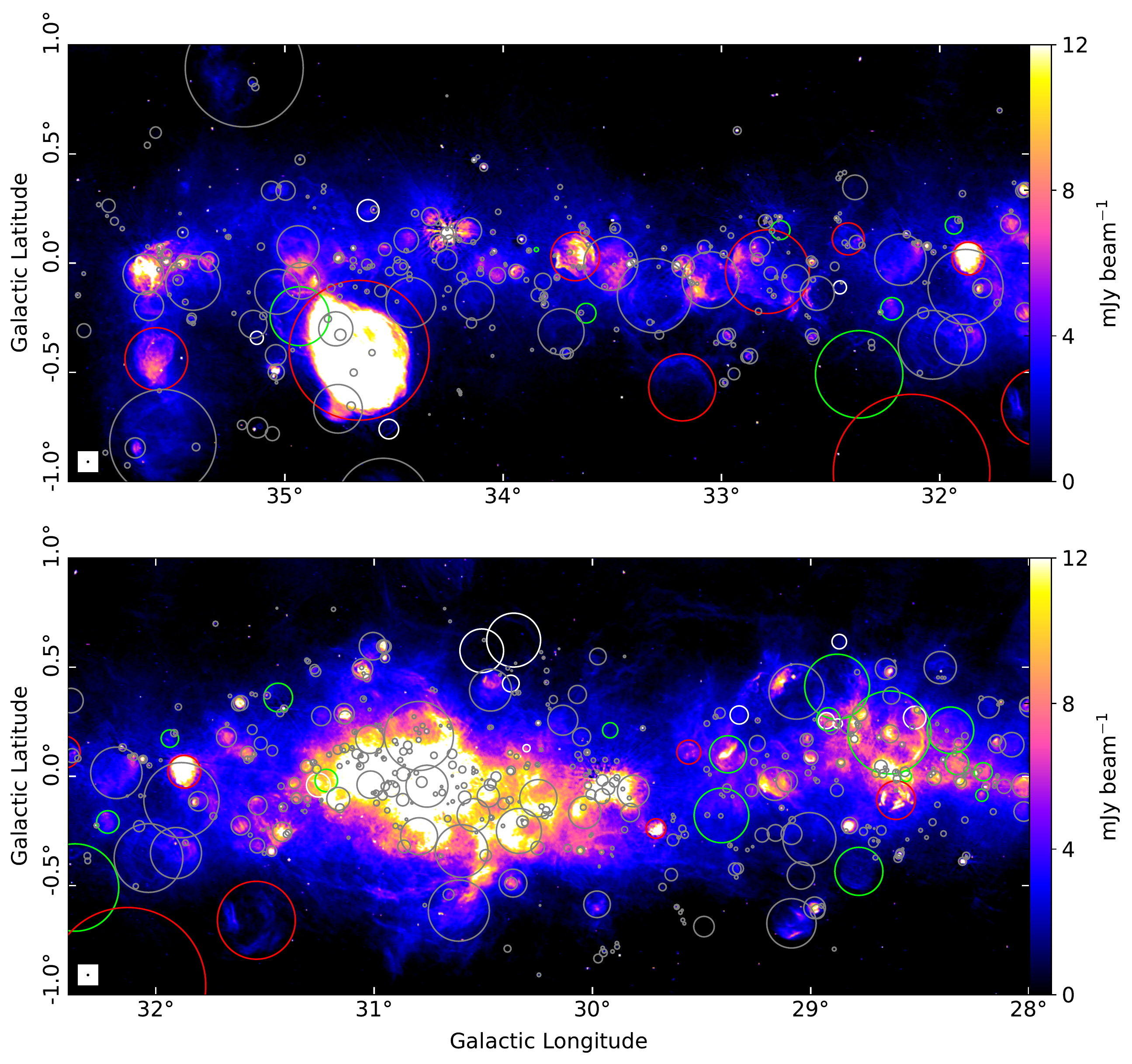}
    \caption{GLOSTAR combination (VLA-D+Effelsberg; \citealt{2021A&A...651A..85B}) image of the region of interest of this study, without the restoration using Urumqi maps (see text for details).  The red, green, and white circles mark the known SNRs \citep{2012AdSpR..49.1313F,2019JApA...40...36G}, the THOR SNR candidates \citep[from][]{2017A&A...605A..58A}, and the GLOSTAR SNR candidates \citep[from][]{2021A&A...651A..86D}, respectively.  The much more numerous H~{\footnotesize{II}} regions, from the WISE catalog \citep{2014ApJS..212....1A} and the GLOSTAR VLA D-configuration catalog \citep{2019A&A...627A.175M}, are marked using grey circles. }
    \label{fig:main}
\end{figure*}

The most recent Galactic SNR catalogs \citep{2012AdSpR..49.1313F,2019JApA...40...36G} contain fewer than 400 objects.  This number is, however, significantly smaller than the expected ${\sim}1000$ discussed by \citet{1991ApJ...378...93L}, who arrived at this estimate by using statistical arguments primarily based on our knowledge of the distances to SNRs in the Milky Way.  \citet{2022arXiv220904570R}, using a similar statistical analysis but with improved distances to the currently known SNRs, estimate that the number quoted by \citet{1991ApJ...378...93L} must be a lower limit, and that there could be over 3000 SNRs in the Galaxy.  It is believed that this apparent discrepancy is only due to SNRs that are fainter and smaller than the currently known sample of SNRs not being detected, rather than insufficient knowledge of the local universe 
(\citealt{2006ApJ...639L..25B,2017A&A...605A..58A}: hereafter \loren).  

In an attempt to improve this situation, recent radio Galactic plane surveys have been carried out with good sensitivity to both compact and extended emission leading to the identification of well over one hundred SNR candidates (\loren; \citealt{2019PASA...36...47H,2021A&A...651A..86D}: \dok ~from here on).  These studies used the radio and mid-infrared (MIR) anti-correlation property of SNRs \citep{1987A&AS...71...63F,1987Natur.327..211H}.  While H~{\footnotesize{II}} regions emit brightly at both radio and MIR wavelengths, SNRs are typically bright radio emitters on the one hand but weak MIR emitters on the other hand.  \citet{1987A&AS...71...63F} found that the ratio ($R$) of $60~\mu$m MIR to 11~cm radio flux density is much higher for H~{\footnotesize{II}} regions than SNRs ($R_{\mathrm{HII}}{\sim}1000$ and $R_{\mathrm{SNR}}{\sim}15$).  Subsequently multiple other studies confirmed this anti-correlation property \citep{1989MNRAS.237..381B,1996A&AS..118..329W,2011AJ....142...47P}.  

Most of these SNR candidates are yet to be confirmed as genuine SNRs with clear nonthermal radio emission.  In addition, some objects in the Galactic SNR catalogs either do not have good radio measurements (such as G32.1-0.9 and G32.4+0.1), or, worse, the evidence that they emit nonthermal synchrotron radiation is rather weak \citep[e.g., G31.5-0.6;][]{2001A&A...370..265M}.  It is not uncommon for H~{\footnotesize{II}} regions, which emit thermally, to be confused as SNRs due to their similar radio morphology (e.g., \loren ~and \dok).  The presence of nonthermal synchrotron radio emission is thus vital to determine whether an object is truly a SNR.  Synchrotron radiation is linearly polarized and typically has a negative spectral index at frequencies where synchrotron losses do not occur \citep[typically over 1~GHz;][]{2013tra..book.....W}, where the spectral index $\alpha$ is determined via  a power law fit to the flux density spectrum, as $S_\nu \propto \nu^\alpha$, $S_\nu$ being the flux density and $\nu$ is the frequency.  In this work, we focus on confirming the status of the SNR candidates and the sample of objects that were catalogued as SNRs (hereafter called `known SNRs') in the region of the Galactic longitude range $28\degree<l<36\degree$ and $|b|<1\degree$ (hereafter called `the pilot region') by measuring linearly polarized flux densities (LPFD) and spectral indices.

The rest of the paper is organized as follows: Section~\ref{sec:data} contains the descriptions of the data and the methods used for this study.  The results for known and candidate SNRs are presented in Sections~\ref{sec:knownSNRs} and \ref{sec:candSNRs} respectively.  Their implications are discussed in Section~\ref{sec:disc}, and we provide a summary of this work in Section~\ref{sec:conclusions}.

\section{Data and methods}
\label{sec:data}

\subsection{The GLOSTAR survey}

The global view on star formation in the Milky Way (GLOSTAR) survey\footnote{\url{https://glostar.mpifr-bonn.mpg.de/glostar/}} is a 4--8~GHz sensitive, unbiased, large scale continuum and spectral line survey of the first quadrant of the Galaxy, covering the region bounded by the Galactic longitudes $-2\degree<l<60\degree$ and Galactic latitudes $|b|<1\degree$, in addition to the Cygnus~X star forming complex ($76\degree<l<83\degree$ and $-1\degree<b<2\degree$).  The observations were done using the Karl Jansky Very Large Array (VLA) in B- and D-configurations, as well as the 100 meter Effelsberg telescope.  
Full details of the observations and data reduction are presented in \citet{2019A&A...627A.175M} and \citet{2021A&A...651A..85B}.  The catalogs of continuum sources in the GLOSTAR pilot region ($28\degree<l<36\degree$ and $|b|<1\degree$) from the VLA images, which contain 1575 sources in the D-configuration images including several supernova remnants and 1457 in the B-configuration images, are discussed in \citet{2019A&A...627A.175M} and \citet{2022arXiv221000560D}, respectively.  An overview of the survey and initial results are described in \citet{2021A&A...651A..85B}.  Further results are presented in \dok~(SNRs), \citet[Cygnus\,X methanol masers]{2021A&A...651A..87O}, \citet[Galactic center continuum sources]{2021A&A...651A..88N}, and \citet[methanol maser catalog]{2022A&A...666A..59N}.  Here, we give a brief overview of the data that we use for this study.

We focus on the GLOSTAR pilot region, which contains numerous extended and compact sources overlapping with a strong Galactic background (see Fig.\,\ref{fig:main} and also \citealt{2019A&A...627A.175M}).  The W43 mini-starburst complex located at $l{\sim}30\degree$ \citep[where the bar of the Milky Way meets the Scutum–Centaurus Arm, e.g.,][]{2014ApJ...781...89Z} and the W44 supernova remnant at $l{\sim}35\degree$ are among the brightest objects observed in this region.  In our previous work (\dok), we detected over 150 SNR candidates in the D-configuration VLA images of the full survey, with the pilot region containing 35 of them.  We use only the D-configuration VLA and the Effelsberg continuum images in this work.

Since the D-configuration images do not completely recover emission on scales larger than a few arcminutes, they are not suitable to accurately measure the total flux densities and spectral indices of extended objects such as many SNRs.  For this purpose, we use images from the single-dish 100\,m Effelsberg telescope and their combination with the VLA D-configuration images.  The Effelsberg images of this survey do not contain information on the very large scales ($> 1\degree$) due to the baseline subtraction and limited coverage in Galactic latitude ($|b|<1\degree$).  The large scale information had been `restored' using the Urumqi 6~cm survey images \citep{2007A&A...463..993S,2011A&A...527A..74S} to produce the GLOSTAR Effelsberg maps with the correct intensities \citep[see][for details]{2021A&A...651A..85B}.  However, all the objects we study are smaller than half a degree, and since we need to filter out the large scale background in any case, we use the original maps directly (i.e., without restoration using the Urumqi maps) to avoid a source of uncertainty.

The calibration and imaging of the VLA and the Effelsberg data, along with their feathering for the total power Stokes~$I$ images, are described by \citet{2021A&A...651A..85B}.  Feathering is a method to combine two images with emission from different angular scales, where the two images are co-added in the Fourier domain ($uv$-space) weighted by their spatial frequency response \citep[e.g.,][]{1984ApJ...283..655V,2017PASP..129i4501C}.  In this work, we exclusively use the combination of VLA-D and Effelsberg data, and hereafter these images are called `the combination images'.
Since the frequency coverage is not exactly the same on both the VLA and the Effelsberg telescopes, producing the combination images is not straightforward.  The final VLA continuum data from the GLOSTAR survey are binned into nine subbands centered on frequencies from ${\sim}4.2$~GHz to ${\sim}7.5$~GHz, whereas the Effelsberg continuum maps have two subbands centered at frequencies $f_{\mathrm{E,lo}}{\sim}$4.9\,GHz and $f_{\mathrm{E,hi}}{\sim}$6.8\,GHz \citep[see][for more details]{2021A&A...651A..85B}.  The procedure that we followed to combine the D-configuration VLA and the Effelsberg data for the different Stokes parameters is described below.

\subsubsection{Image combination: Stokes~$I$}
\label{subsec:comb_I}

We average the VLA images from the first five subbands at lower frequencies and the next three subbands at higher frequencies separately to form two VLA images, $I_{\mathrm{V,f_{\mathrm{V,lo}}}}$ and $I_{\mathrm{V,f_{\mathrm{V,hi}}}}$ respectively.  We discard the ninth subband since it is mostly corrupted by radio frequency interference.  The first five subbands have an average frequency $f_{\mathrm{V,lo}}{\sim}$4.7\,GHz and the next three subbands have $f_{\mathrm{V,hi}}{\sim}$6.9\,GHz, which are already close to the central frequencies of the Effelsberg continuum data, $f_{\mathrm{E,lo}}{\sim}$4.9\,GHz and $f_{\mathrm{E,hi}}{\sim}$6.8\,GHz respectively, but they are not exactly equal.  To bring the two VLA images ($I_{\mathrm{V,f_{\mathrm{V,lo}}}}$ and $I_{\mathrm{V,f_{\mathrm{V,hi}}}}$) to the frequencies of the Effelsberg images, we use a pixel-by-pixel VLA spectral index $\alpha_\mathrm{pix}$ to scale the intensities of each pixel:
\begin{equation}
    I_{\mathrm{V,f_{\mathrm{E,lo}}}}^{\mathrm{pix}} = I_{\mathrm{V,f_{\mathrm{V,lo}}}}^{\mathrm{pix}} \left(\frac{f_{\mathrm{E,lo}}}{f_{\mathrm{V,lo}}}\right)^{\alpha_\mathrm{pix}} ~\mathrm{and}~~ I_{\mathrm{V,f_{\mathrm{E,hi}}}}^{\mathrm{pix}} = I_{\mathrm{V,f_{\mathrm{V,hi}}}}^{\mathrm{pix}} \left(\frac{f_{\mathrm{E,hi}}}{f_{\mathrm{V,hi}}}\right)^{\alpha_\mathrm{pix}}
\end{equation}
where $I_{\mathrm{V,f_{\mathrm{E,lo}}}}^{\mathrm{pix}}$ and $I_{\mathrm{V,f_{\mathrm{E,hi}}}}^{\mathrm{pix}}$ are the VLA pixel values estimated at the Effelsberg central frequencies.  For pixels with intensities above a signal-to-noise threshold of three, $\alpha_\mathrm{pix}$ is measured from the two Stokes~$I$ images $I_{\mathrm{V,f_{\mathrm{V,lo}}}}$ and $I_{\mathrm{V,f_{\mathrm{V,hi}}}}$.  For pixels below the threshold, we take a spectral index of zero, i.e., we use the average intensity.  After bringing the VLA images to the central frequencies of the Effelsberg images, we feather the VLA and the Effelsberg maps $I_{\mathrm{V,f_{\mathrm{E,lo}}}} + I_{\mathrm{E,f_{\mathrm{E,lo}}}}$ to produce the low frequency combination image, and $I_{\mathrm{V,f_{\mathrm{E,hi}}}} + I_{\mathrm{E,f_{\mathrm{E,hi}}}}$ to produce the high frequency combination image.  Finally, the low and high frequency images are averaged to form the 5.85~GHz GLOSTAR combination image.  These combination images will be made available on the GLOSTAR image server\footnote{\url{https://glostar.mpifr-bonn.mpg.de/glostar/image\_server}} before the publication of this work.

\subsubsection{Image combination: Stokes~$Q$ and $U$}

Similar to the Stokes~$I$ procedure, we make the low- and high-frequency VLA images by averaging the first five and next three subbands.  We then directly feather each of these averaged images with their respective Effelsberg maps: $P_{\mathrm{V,f_{\mathrm{V,lo}}}} + P_{\mathrm{E,f_{\mathrm{E,lo}}}}$ and $P_{\mathrm{V,f_{\mathrm{V,hi}}}} + P_{\mathrm{E,f_{\mathrm{E,hi}}}}$, where $P$ represents Stokes~$Q$ or $U$.  We do this without any intensity scaling applied to bring them to the exact same frequency as we did for Stokes~$I$.  This is because Stokes $Q$ and $U$ have both positive and negative features, and a direct spectral index calculation is not possible.  The Stokes $Q$ and $U$ images at each of the two frequencies are then combined to form the linearly polarized intensity maps $\sqrt{Q^2+U^2}$.  The low and high frequency maps are then averaged to form the 5.85~GHz GLOSTAR combination image of linearly polarized intensity.

We note that this method may introduce a bias in the measured polarized intensities and the polarization vectors due to the different central frequencies.  However, we find that this bias is negligible since the frequencies are quite close ($f_{\mathrm{V,lo}} \approx f_{\mathrm{E,lo}}$ and $f_{\mathrm{V,hi}} \approx f_{\mathrm{E,hi}}$).  Assuming a spectral index of $-0.7$ for synchrotron emission, the different central frequencies of the feathered VLA and Effelsberg images of linearly polarized emission introduce an error of approximately $4\%$, which is close to the calibration uncertainty.  For the polarization vector to change by just five degrees from $f_{\mathrm{V,lo}}$ to $f_{\mathrm{E,lo}}$, the rotation measure must be greater than about $2500~\mathrm{rad~m}^{-2}$, which is unlikely to be seen in any typical Galactic source.  Nonetheless, to introduce this bias in the uncertainty measurement of flux densities and also the instrumental polarization (${\lesssim}2\%$ in both VLA and Effelsberg data), we adopt a conservative $10\%$ error that will be added in quadrature to the usual uncertainty we obtain from the measurement of flux density of an extended source.  In addition, we observe that the LPFD measured in the combination images may be lower than the values measured in the VLA D-configuration only images.  This can happen due to the depolarization that occurs when the polarization vectors in the small scale structure detected by the VLA are misaligned with the polarization vectors measured from the Effelsberg data.  It is worth noting that, in this study, the exact degree of polarization is not exceptionally important except to the degree it establishes whether the source is or is not polarized, i.e., we only use it as a tool to identify nonthermal emission.

\subsection{Supernova remnant catalogs}

In \dok, we presented the list of the SNR candidates that are detected in the D-configuration VLA images of the GLOSTAR survey.  It contains 77 objects that were noted as potential SNRs by earlier studies (their table~3), and 80 new identifications as well (their table~4).  These candidates were identified using the MIR-radio anti-correlation property of SNRs as discussed earlier.  

In the GLOSTAR pilot region, there are 21 candidates discovered in the 1--2 GHz HI/OH/Recombination line survey (THOR; \loren) and 14 from the GLOSTAR survey (\dok).  These 35 candidates, in addition to the 12 confirmed SNRs in the Galactic SNR catalogs by \citet{2012AdSpR..49.1313F} and \citet{2019JApA...40...36G}, are the targets of this study.

\subsection{Ancillary data}
\label{subsec:ancillary}

In addition to the GLOSTAR survey continuum images previously described, we also use other complementary radio surveys that are able to recover emissions at the scale of several arcminutes: the 1--2 GHz HI/OH/Recombination line survey \citep{2016A&A...595A..32B,2020A&A...634A..83W} combined with the VLA Galactic Plane Survey \citep[VGPS;][]{2006AJ....132.1158S}, which is called the THOR+VGPS\footnote{\url{https://www2.mpia-hd.mpg.de/thor/Data\_\%26\_Publications.html}}, the 80–230 MHz GaLactic and Extragalactic All-sky Murchison Widefield Array survey \citep[GLEAM;][]{2019PASA...36...47H}\footnote{\url{http://gleam-vo.icrar.org/gleam\_postage/q/form}}, the Effelsberg 11 cm (${\sim}2.7$ GHz) survey of the Galactic plane by \citet{1984A&AS...58..197R}\footnote{\url{https://www3.mpifr-bonn.mpg.de/survey.html}}, and the 3~cm (10~GHz) survey of the Galactic plane with the Nobeyama telescope by \citet{1987PASJ...39..709H}\footnote{\url{http://milkyway.sci.kagoshima-u.ac.jp/~handa/}}.

\subsection{Flux density and spectral index measurements}
\label{subsec:fluxdspidxmeas}

We use the GLOSTAR combination images to measure the flux densities at 5.85~GHz, in addition to other surveys as mentioned earlier (Section\,\ref{subsec:ancillary}).  We note that we do not measure the flux densities from the two sub-bands to derive an `in-band' GLOSTAR spectral index, since each of those images depends upon---though only partly---the pixel-by-pixel spectral index from the VLA data, which suffer from the problem of the undetected large scale flux density (see Section\,\ref{subsec:comb_I}).  

The presence of background emission may bias the value of the measured spectral index.  This is particularly true for extended objects in the Galactic plane since the nonthermal Galactic background is strong and ubiquitous at low radio frequencies.  In addition, the intensity of this background is dependent on frequency and position \citep[e.g.,][]{2005MNRAS.360.1545P}.  The method of `unsharp masking' \citep{1979A&AS...38..251S} is generally used to filter out the large scale Galactic emission, but it is not appropriate for smaller scale background emission across an object with the size of a few arcminutes.  In this work, we fit a `twisted plane' that removes the background contribution up to a first order variation.  Points are chosen around an object such that they represent the background emission in that area, and a two-dimensional least-squares linear fit is performed to the pixel intensities to measure the background variation.  The uncertainty from this background subtraction operation is determined by choosing multiple sets of vertices.  We subtract the local background in both the total intensity and the polarized intensity images, and we mask pixels typically below a $3\sigma$-level, where the noise is determined locally by a sigma-clipping algorithm.  

While several objects we discuss in this paper already have their low frequency flux densities derived in multiple previous studies, for the sake of consistency with regards to spectral index, we make our own measurements of the flux densities of these objects using the images directly from their survey data, performing background subtraction in the same manner as we do for the GLOSTAR images.  We also mask the point sources that are clearly unrelated \citep[e.g.,][]{2005A&A...436..187T} to keep the measurement as accurate as possible.  In addition, since radio interferometric artefacts such as radial `spokes' are common near bright sources, we do not measure the total or linearly polarized flux densities if we are unable to disentangle such effects from the emission of an object.  Due to such artefacts, polarization measurements are not possible for about a third of the objects studied in this work.

The spectral index of an object is usually measured by fitting a straight line to the relation between flux densities and frequencies in logarithmic space:
\begin{equation}
    \alpha_\mathrm{FD} = \frac{\log(S_\nu)}{\log(\nu)}
\end{equation}
However, the values determined in this manner are sensitive to the presence of background emission.  \citet{1962MNRAS.124..297T} introduced the concept of temperature-temperature (TT) plots, in which a spectral index is extracted from the slope of a straight line fit to the pixel intensities at one frequency against the pixel intensities at another frequency.  In essence,  we integrate over the whole area to measure the flux density spectral index ($\alpha_\mathrm{FD}$), whereas the TT-plot spectral index ($\alpha_\mathrm{TT}$) is calculated by measuring the variation of each pixel at different frequencies.  

The intensities on TT-plots can be represented by brightness temperatures in Kelvin, or pixel intensities in $\mathrm{Jy~beam}^{-1}$.  In this work, we exclusively use pixel intensities, and the spectral index is calculated using:
\begin{equation}
    \alpha_\mathrm{TT} = \frac{\log(m_\mathrm{S})}{\log(\nu_1/\nu_2)}
\end{equation}
where $m_\mathrm{S}$ is the slope of the line that is fit to pixel intensities.  This is a more reliable measurement of spectral index of an extended object because the flux density bias introduced by a constant large scale background emission moves all the points equally, and hence does not affect the slope of the fit.  Since the combination images are produced using the spectral index derived from the D-configuration GLOSTAR-VLA images, they are not suitable to measure the TT-plot spectral index ($\alpha_\mathrm{TT}$).  We only use the GLOSTAR-Effelsberg images for this purpose.  We also measure the flux density spectral index ($\alpha_\mathrm{FD}$); this serves as a useful consistency check since we subtract the background regardless, as described above.

We note that, at low radio frequencies such as the regime of the GLEAM survey ($\lesssim$200~MHz), absorption effects become important, either via synchrotron self-absorption or free-free absorption \citep[e.g.,][]{2013tra..book.....W,2019AJ....158..253A}.  This lowers the emitted flux at low frequencies and increases the power-law spectral index compared to values determined at higher frequencies.  Such a `spectral break' effect had been noted in several SNRs before \citep[e.g.,][]{2011A&A...536A..83S}.  Spectral breaks are also observed in pulsar wind nebulae due to the central pulsar's time-dependent energy injection \citep{1984ApJ...278..630R,1997A&A...325..295W}.  When we calculate the flux density spectral index in this work, if we clearly see a break, we split the spectrum into two and calculate two spectral indices; if the break is not obvious, we calculate only a single spectral index.
\begin{table*}[!h]
\centering
\caption{Flux densities and spectral indices of the known SNRs in the pilot region.}
\label{tab:knownSNRs}
\begin{tabular}{l c c c c c c c}
\hline\hline
    Name  & $S_{\mathrm{0.2GHz}}$ & $S_{\mathrm{1.4GHz}}$ & $S_{\mathrm{2.7GHz}}$ & $S_{\mathrm{5.8GHz}}$ & $p_{\mathrm{5.8GHz}}$ & $S_{\mathrm{10.5GHz}}$ & $\alpha_\mathrm{FD}$  \\
                 & (Jy)                & (Jy)                & (Jy)                & (Jy)                & (\%)                 & (Jy)             &        \\
\hline
    G28.6$-$0.1  &  $ 15.2 \pm 1.6 $   &  $ 4.4 \pm 0.4 $    &  $ 3.4 \pm 0.4 $    &  $ 1.7 \pm 0.2 $    &   $ 2.6 \pm 0.6 $   &  $ 1.4 \pm 0.2 $  &  $ -0.61 \pm 0.04 $  \\
    G29.6$+$0.1  &  $ 0.84 \pm 0.11 $  &  $ 0.48 \pm 0.06 $  &  $ 0.25 \pm 0.03 $  &  $ 0.14 \pm 0.02 $  &        $ < 71 $     &  $ - $            &  $ -0.50 \pm 0.12 $  \\
    G29.7$-$0.3  &  $ 25.4 \pm 2.6 $   &  $ 7.0 \pm 0.7 $    &  $ 3.9 \pm 0.4 $    &  $ 3.1 \pm 0.3 $    &   $ 2.3 \pm 0.8 $   &  $ 1.5 \pm 0.2 $  &  $ -0.69 \pm 0.05 $  \\
    G31.5$-$0.6  &  $ 1.9 \pm 0.6 $    &  $ 1.6 \pm 0.5 $    &  $ 1.7 \pm 0.2 $    &  $ 1.6 \pm 0.2 $    &        $ < 81 $     &  $ 1.5 \pm 0.2 $  &  $ -0.04 \pm 0.09 $  \\
    G31.9$+$0.0  &  $ 34.3 \pm 3.4 $   &  $ 16.3 \pm 1.6 $   &  $ 12.8 \pm 1.3 $   &  $ 8.8 \pm 0.9 $    &   $ 0.8 \pm 0.2 $   &  $ 6.2 \pm 0.7 $  &  $ -0.42 \pm 0.03 $  \\
    G32.1$-$0.9  &  $ 12.5 \pm 2.4 $   &  $ 3.5 \pm 0.5 $    &  $ 2.1 \pm 0.5 $    &  $ - $              &         $ - $       &  $ - $            &  $ -0.68 \pm 0.11 $  \\
    G32.4$+$0.1  &  $ 1.1 \pm 0.2 $    &  $ 0.75 \pm 0.13 $  &  $ - $              &  $ 0.53 \pm 0.07 $  &       $ < 58 $      &  $ - $            &  $ -0.21 \pm 0.07 $  \\
    G32.8$-$0.1  &  $ 16.3 \pm 1.7 $   &  $ 10.9 \pm 1.1 $   &  $ 7.3 \pm 0.8 $    &  $ 6.5 \pm 0.9 $    &   $ 2.5 \pm 1.3 $   &  $ 5.9 \pm 0.7 $  &  $ -0.27 \pm 0.04 $  \\
    G33.2$-$0.6  &  $ 5.0 \pm 0.5 $    &  $ 2.9 \pm 0.3 $    &  $ 2.3 \pm 0.3 $    &  $ 1.9 \pm 0.3 $    &   $ 0.11 \pm 0.07 $ &  $ - $            &  $ -0.29 \pm 0.05 $  \\
    G33.6$+$0.1  &  $ 26.6 \pm 2.7 $   &  $ 10.9 \pm 1.1 $   &  $ 6.9 \pm 0.8 $    &  $ 5.5 \pm 0.7 $    &   $ 8.1 \pm 1.9 $   &  $ 3.5 \pm 0.5 $  &  $ -0.50 \pm 0.04 $  \\
    G34.7$-$0.4  &  $ 320 \pm 32 $     &  $ 193 \pm 19 $     &  $ 145 \pm 15 $     &  $ 112 \pm 11 $     &     $ 6.8 \pm 1.4 $ &  $ 57 \pm 6 $     &  $ -0.40 \pm 0.07 $  \\
    G35.6$-$0.4  &  $ 8.6 \pm 0.9 $    &  $ 8.8 \pm 0.9 $    &  $ 7.5 \pm 0.8 $    &  $ 6.0 \pm 0.6 $    &   $ 7.1 \pm 2.0 $   &  $ 4.7 \pm 0.5 $  &  $ +0.02 \pm 0.08^\dagger $  \\
                 &                     &                     &                     &                     &                     &                   & $ -0.31 \pm 0.07^\dagger $   \\
\hline
\end{tabular}
\tablefoot{ $S_{\mathrm{0.2GHz}}$, $S_{\mathrm{1.4GHz}}$, $S_{\mathrm{2.7GHz}}$, $S_{\mathrm{5.8GHz}}$, and $S_{\mathrm{10.5GHz}}$ are the continuum flux densities we measured from the 200~MHz GLEAM, the 1.4~GHz THOR+VGPS, 11~cm Effelsberg, the 5.8~GHz GLOSTAR combination, and the 3~cm Nobeyama survey images.  $p_{\mathrm{5.8GHz}}$ is the percentage linear polarization measured in the GLOSTAR combination images.  If no emission is found, then the $1\sigma$ upper limits are quoted.  $\alpha_\mathrm{FD}$ is the broadband spectral index derived from the measured flux densities.\\
$^\dagger$ Since a break in the spectrum is clearly visible, we report both the spectral indices.
}
\end{table*}

The nonthermal emission from the Galactic disk is polarized, and it may have structure on small scales that is not filtered out by an interferometer.  While this is more significant at longer wavelengths, it might affect the GLOSTAR images too (see \dok).  We verify that in the objects we study in this work, there exist no features with no Stokes~$I$ counterparts when measuring LPFD.  In addition, a Ricean polarization bias might introduce a positive offset.  This occurs because LPFD is the square root of the sum of squares ($\sqrt{Q^2+U^2}$), and any positive or negative noise in $Q$ and $U$ will always add up and result in a non-zero LPFD.  We find that this effect is an order of magnitude smaller than the flux densities we report, and in fact there is no need to correct for this bias due to the background subtraction procedure and the $3\sigma$-level mask we use \citep[see][]{1974ApJ...194..249W}.  Nonetheless, the twisted-plane background subtraction procedure is applied to the linearly polarized intensity images as well, which accounts for the Galactic plane polarized background and also any possible Ricean polarization bias.

\section{Known SNRs}
\label{sec:knownSNRs}

\begin{figure*}[h!]
    \centering
    \includegraphics[width=\textwidth]{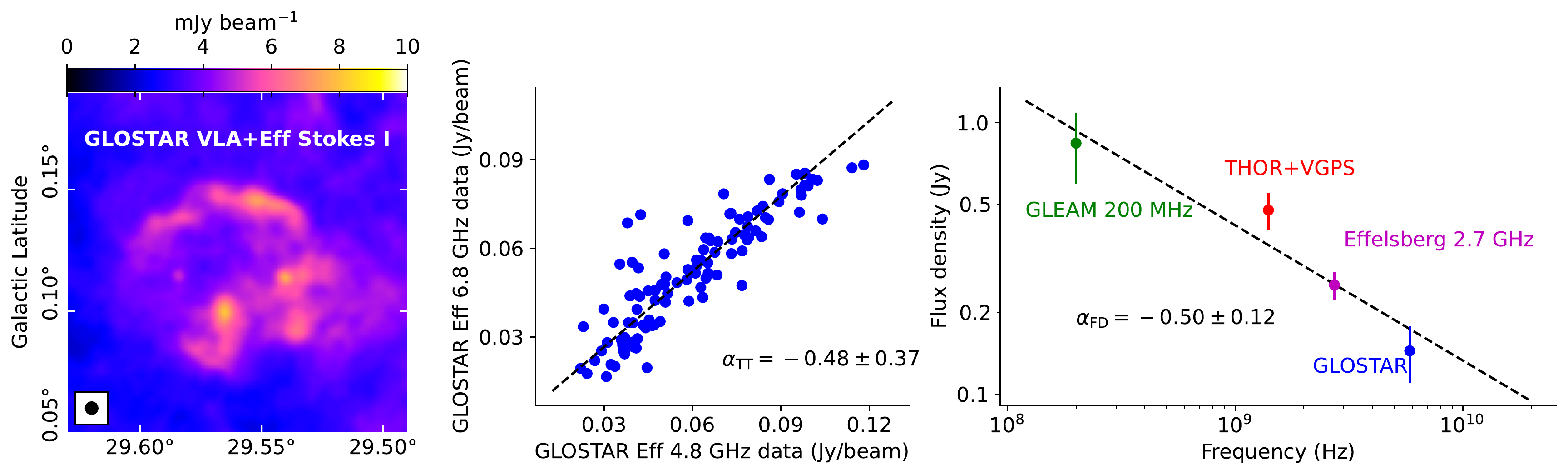}\\
    \includegraphics[width=\textwidth]{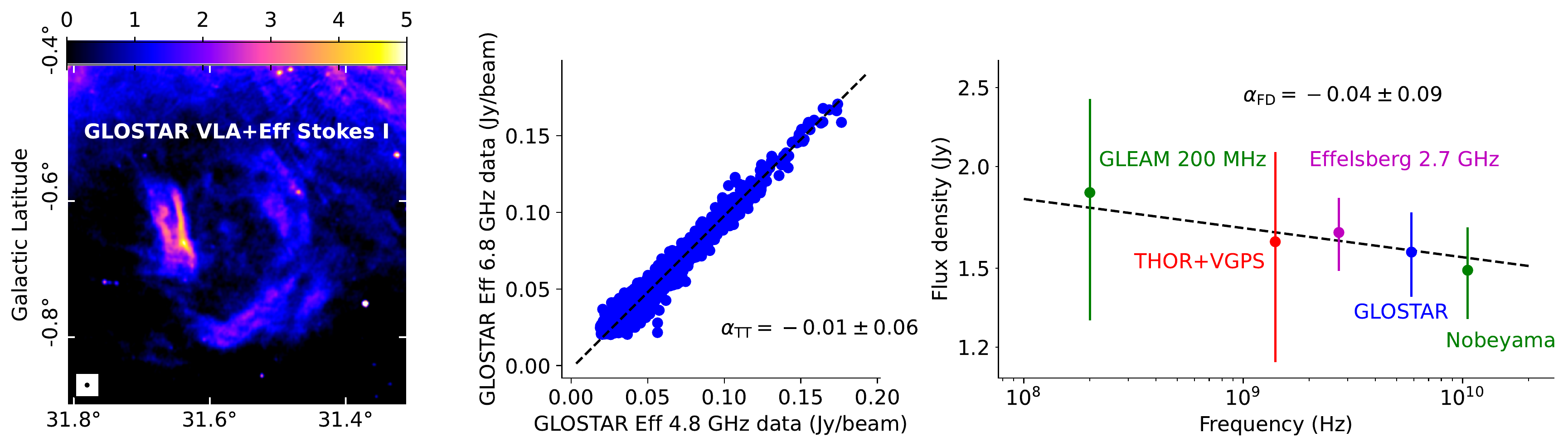}\\
    \includegraphics[width=\textwidth]{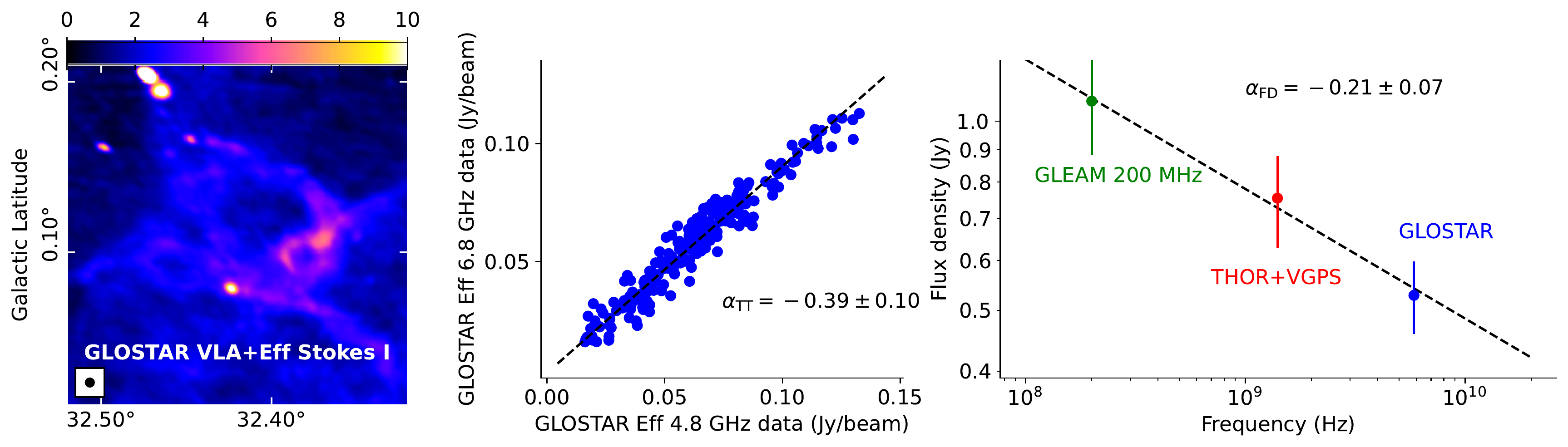}\\
    \includegraphics[width=\textwidth]{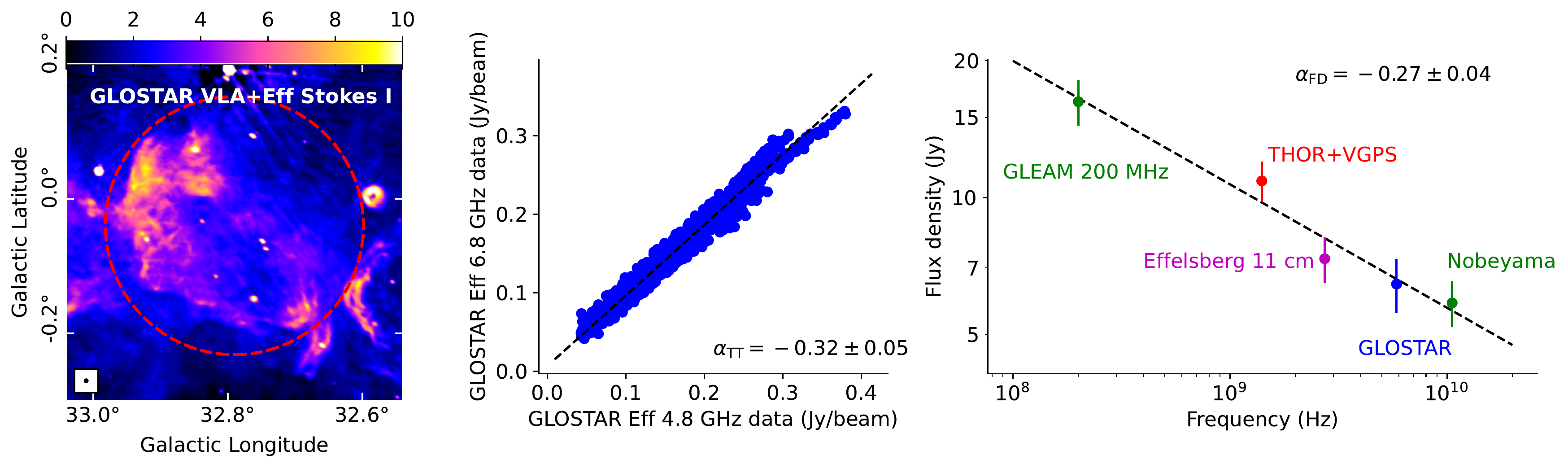}
    \caption{From top to bottom: G29.6+0.1, G31.5-0.6, G32.4+0.1, and G32.8-0.1.  Left panels show the GLOSTAR combination images.  The TT-plot from GLOSTAR-Effelsberg images, and the flux density spectrum using the GLOSTAR combination images and ancillary data are presented in the middle and right panels respectively.  }
    \label{fig:ksnrs}
\end{figure*}

The Galactic SNR catalogs of \citet{2019JApA...40...36G} and \citet{2012AdSpR..49.1313F} list 12 confirmed SNRs in the region we study.  In Table~\ref{tab:knownSNRs}, we present the GLOSTAR 5.8~GHz integrated flux densities of these SNRs along with their spectral indices.  If possible, overlapping H~{\footnotesize{II}} regions and clearly unrelated point sources are masked while measuring the flux densities, taking special care in crowded regions.  If it is unclear whether a particular region of emission belongs to the SNR, we do not remove that region.  We find that the flux densities and spectral indices are generally consistent with previous studies.  We present the GLOSTAR combination images of some interesting known SNRs and discuss them below.  The total intensity images and the linearly polarized intensity images of all the known SNRs studied in this work are shown in Appendix~\ref{apdx:knownSNRs}.

\subsection{G29.6+0.1}

While we had already detected linear polarization in the SNR G29.6+0.1 using the VLA D-configuration images (in \dok), the emission in the combination images seems to be depolarized due to the addition of large scale information from the GLOSTAR-Effelsberg data.  We do not measure any polarized emission over a $1\sigma$ upper limit of ${\sim}0.1$~Jy.  The flux densities we measure (see Table~\ref{tab:knownSNRs}) appear to be lower than what is expected from the lower limits reported by \citet{1999ApJ...526L..37G}: ${\sim}0.41$~Jy and ${\sim}0.26$~Jy at 5~GHz and 8~GHz, respectively.  The reason for this inconsistency is unclear.  Nonetheless, the broadband spectral index we derive from our measurements (${\sim}-0.5$) is in line with the TT-plot spectral indices derived by \citet{1999ApJ...526L..37G}.  We show the GLOSTAR combination image of the SNR G29.6+0.1 in Fig.~\ref{fig:ksnrs}.  The spectrum of this SNR shows that it might be falling more rapidly from 1.4--5.8~GHz than from 0.2--1.4~GHz, suggesting the presence of a spectral break around 1~GHz.  But given the uncertainties, we reserve judgment on the changing spectral index.

\subsection{G31.5-0.6}

We show the GLOSTAR combination Stokes~$I$ image of the known SNR G31.5-0.6 along with its flux density spectrum in Fig.~\ref{fig:ksnrs}.  We find no significant linear polarization in agreement with the observations of \citet{1987A&AS...69..403F}, who suggest that this is a SNR--H~{\footnotesize{II}} region complex.  The Stokes~$I$ flux densities we measure are consistent with those given by \citet{1987A&AS...69..403F} within uncertainties, and we also find a morphology similar to their image.  However, the spectral index we derive from 200~MHz to 10~GHz is essentially zero, which is consistent with our TT-plot result (Fig.~\ref{fig:ksnrs}), but in slight tension with the value of ${\sim}-0.2$ given by \citet{1987A&AS...69..403F}.  Even after separating from the region the thermal emission that they reported, we find no evidence for synchrotron emission.  In the $24~\mu$m images of MIPSGAL \citep{2009PASP..121...76C}, we find weak emission following the radio morphology, hinting that the emission may be thermal.  Based on sulfur and H$\alpha$ optical lines, \citet{2001A&A...370..265M} also suggest that this may be an H~{\footnotesize{II}} region instead of a SNR.  High resolution deeper observations at lower frequencies will shed more light on the nature of the emission from this object, but the evidence so far suggests that G31.5-0.6 is not a SNR.

\subsection{G32.1-0.9}

\begin{figure*}[h!]
    \centering
    \includegraphics[width=0.8\textwidth,trim={0 0 0 -15mm}, clip]{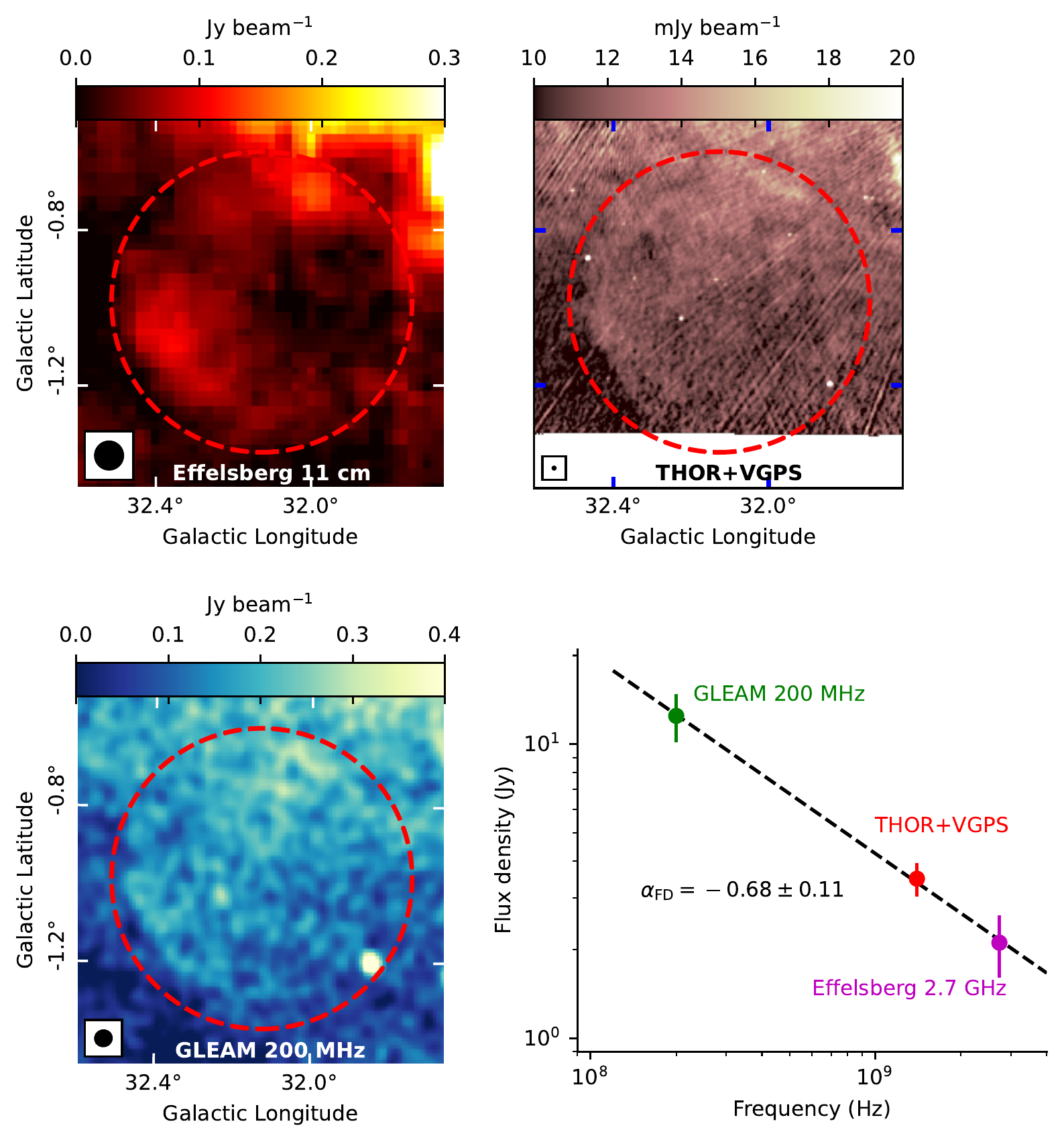}
    \caption{SNR G32.1-0.9 as seen in the Effelsberg 11~cm (top left), THOR+VGPS (top right), and the GLEAM 200~MHz images (bottom left).  Its flux density spectrum is shown in the bottom right panel.}
    \label{fig:G32.1-0.9}
\end{figure*}

\citet{1997MNRAS.292..365F} discovered the SNR G32.1-0.9 in the X-ray regime, and they found a possible faint radio counterpart in the 11~cm Effelsberg images.  \loren ~reported a possible detection in the THOR+VGPS data too, but no radio spectral index was ever determined.  While we cannot confidently identify any counterpart in the GLOSTAR data, the 200~MHz GLEAM image shows a shell that corresponds to the 11~cm Effelsberg and THOR+VGPS detections (Fig.~\ref{fig:G32.1-0.9}).  Using these three images, we derive a radio spectral index for this unusually faint SNR for the first time, $ \alpha_\mathrm{FD} = -0.68 \pm 0.12 $.  Its average 1~GHz surface brightness is approximately $ 3~\times~10^{-22}~\mathrm{W}~\mathrm{m}^{-2}~\mathrm{Hz}^{-1}~\mathrm{sr}^{-1} $, which makes it one of the faintest radio SNRs currently known: it is only three times brighter than the faintest SNR known in the Milky Way \citep[G181.1+9.5;][]{2017A&A...597A.116K}.

\subsection{G32.4+0.1}

G32.4+0.1 was discovered in the X-ray regime by \citet{2004PASJ...56.1059Y}, who also noted a possible counterpart in the images of the 1.4~GHz NRAO VLA Sky Survey \citep{1998AJ....115.1693C}.  The radio emission from this SNR is faint but clearly visible in the GLEAM, the THOR+VGPS and the GLOSTAR combination images, allowing us to measure, for the first time for this SNR, a spectral index of $-0.21\pm 0.07$ (from brightness values) to $-0.39\pm 0.10$ (from a TT-plot).  The GLOSTAR combination image and the plots for spectral index determination are shown in Fig.~\ref{fig:ksnrs}.  As noted in \S\ref{subsec:fluxdspidxmeas}, the low frequency emission detected in GLEAM may be self-absorbed which brings the spectral index close to zero; hence we favor the TT-plot spectral index (${\sim}-0.4$) for higher frequencies where the effects of synchrotron self-absorption are not important.  Linear polarization is undetected, with an upper limit on the linearly polarized flux density of ${\sim}0.3$~Jy.

\subsection{G32.8-0.1}

\begin{figure*}[h]
    \centering
    \includegraphics[width=16cm]{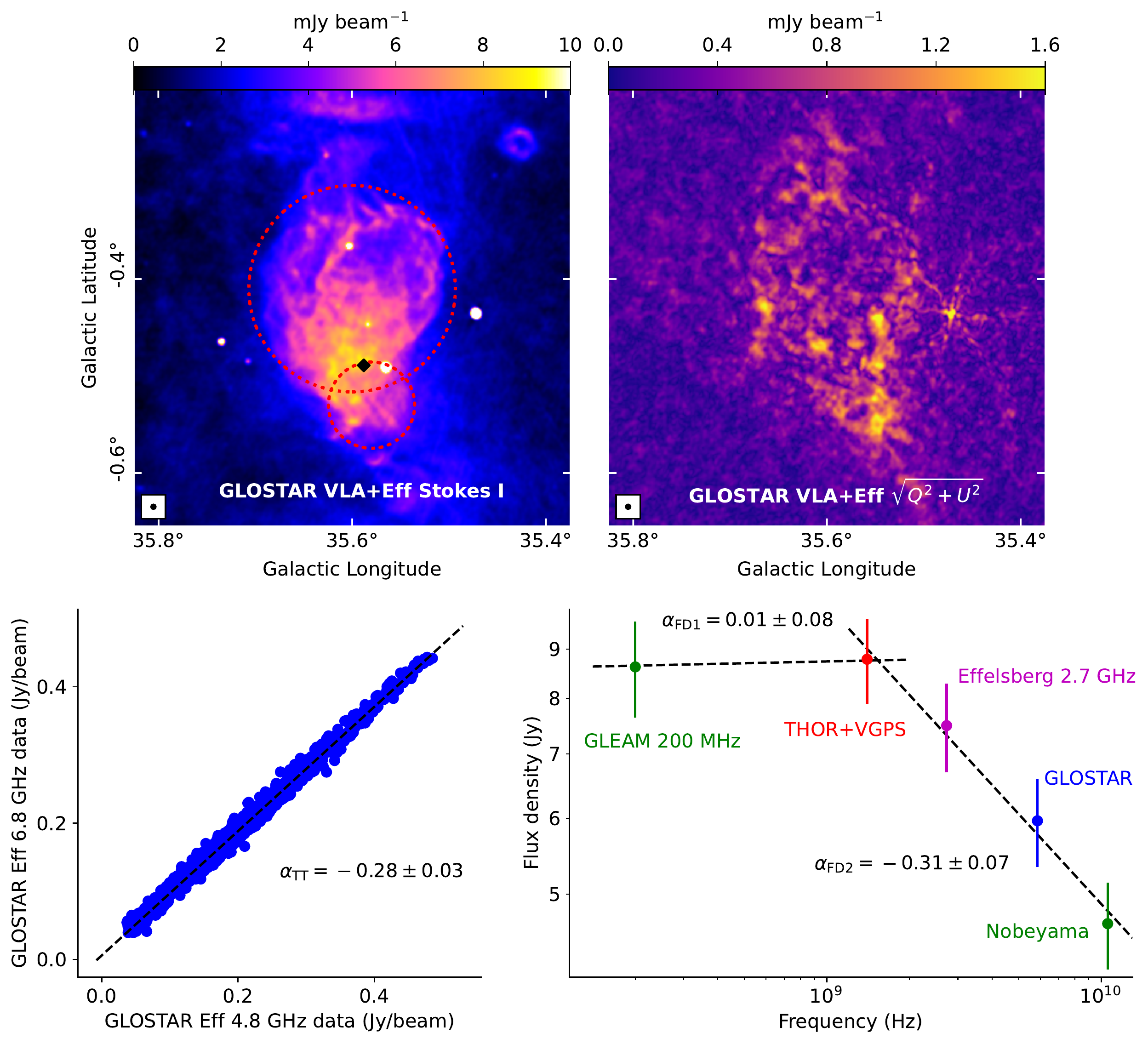}
    \caption{SNR G35.6-0.4: the top left and right panels show the GLOSTAR combination images of total and linearly polarized intensity.  The red dotted circles represent the two shell-like structures identified at 610~MHz by \citet{2014A&A...561A..56P}, while the black diamond marks the position of the recombination line discovered by \citet{1989ApJS...71..469L}.  The TT-plot from GLOSTAR-Effelsberg images and the flux density spectrum are presented in the bottom left and right panels respectively. }
    \label{fig:G35.6-0.4}
\end{figure*}

\citet{2019JApA...40...36G} lists the SNR G32.8-0.1 with an uncertain spectral index of $-0.2$ based on the work of \citet{1975AuJPA..37...39C}, who report a flux density of 12.8~Jy at 408~MHz.  Unfortunately, no uncertainties were quoted, but they reported that their error might be large.  Later, \citet{1992AJ....103..943K} observed this SNR with the VLA at a similar frequency of 330~MHz, and their results are in dispute with the result from \citet{1975AuJPA..37...39C}.  They measured a significantly higher flux density of ${\sim}32$~Jy and consequently a more negative spectral index of ${\sim}-0.5$, but no uncertainties were quoted once again.  This SNR is clearly visible in the GLOSTAR survey, in addition to the GLEAM and the THOR surveys (Fig.~\ref{fig:ksnrs}), which helps us resolve the tension.  Our measurements of flux density ($16.3\pm 1.7$~Jy at 200\,MHz) and spectral index ($\alpha_{\mathrm{TT}}=-0.27\pm 0.04$) are consistent with the values given by \citet{1975AuJPA..37...39C}, which is confirmed by our TT-plot spectral index ($\alpha_{\mathrm{TT}}=-0.32\pm 0.05$).

\subsection{G35.6-0.4}

The nature of emission from G35.6-0.4 had been a subject of discussion for a long time.  It was included in the early SNR catalogues \citep[e.g.,][]{1971AJ.....76..305D,1979AuJPh..32...83M}, but the detection of a radio recombination line by \citet{1989ApJS...71..469L} among other studies, had cast doubts that the emission is nonthermal \citep[see][for an overview]{2009MNRAS.399..177G}.  Finally, using higher quality radio continuum data, \citet{2009MNRAS.399..177G} ``re-identified'' this as a SNR with a spectral index of ${\sim}-0.5$.  This object is clearly visible in the GLOSTAR survey, where we also unambiguously detect linearly polarized emission (Fig.~\ref{fig:G35.6-0.4}).  Its spectrum appears to be broken; from 200~MHz to 1.4~GHz the flux density has no significant change ($\alpha{\sim}0$), and from 1.4~GHz to 10~GHz it falls with a spectral index of $\alpha=-0.31\pm 0.07$.  This is confirmed with the GLOSTAR Effelsberg TT-plot spectral index as well (Fig.~\ref{fig:G35.6-0.4}).  This result ($\alpha{\sim}-0.3$) is also quite consistent with the spectral index derived by \citet{2022ApJ...933..187R}: $\alpha=-0.34\pm 0.08$ from 2.7--30~GHz. \citet{2009MNRAS.399..177G} derives a slightly steeper spectral index of $-0.47\pm 0.07$.  This is probably because of the different choice of polygons used for measuring the flux density and also the subtraction of background emission in this complex region, but we note that the values are consistent within $2\sigma$.  

Given the presence of radio recombination lines that indicate thermal emission and a spectral index ${\sim}-0.3$, this region appears to be a complex of thermal and nonthermal emissions.  \citet{2014A&A...561A..56P} suggest that there may be two circularly shaped extended objects present in this complex (marked by two red dotted circles in Fig.~\ref{fig:G35.6-0.4}), and with one of them with thermal and the other one with nonthermal emission.  We find that MIR emission is detected from the southern part in the GLIMPSE and MIPSGAL images \citep{2009PASP..121..213C,2009PASP..121...76C}, providing further evidence of thermal emission from this region.  The linearly polarized emission detected in the GLOSTAR combination data (see Fig.~\ref{fig:G35.6-0.4}) also hints at the presence of two shells, one centered at G35.60-0.40 and the other at G35.55-0.55, similar to those reported by \citet{2014A&A...561A..56P}.  However, since we find polarization from both these regions, it is likely that emissions from these regions have both thermal and nonthermal components.

\section{Candidate SNRs}
\label{sec:candSNRs}

\begin{table*}
\centering
\caption{Flux densities and spectral indices of the THOR SNR candidates in the pilot region.}
\label{tab:THORcands}
\begin{tabular}{l c c c c c l}
\hline\hline
Name & $S_{\mathrm{0.2GHz}}$ & $S_{\mathrm{1.4GHz}}$ & $S_{\mathrm{5.8GHz}}$ & $p_{\mathrm{5.8GHz}}$ & $\alpha_\mathrm{FD}$ & Remarks \\
& (Jy) & (Jy) & (Jy) & (\%) & & \\
\hline
    G28.21$+$0.02  &  $ - $             &  $ 0.27 \pm 0.03 $  &  $ 0.18 \pm 0.02 $  &  $ - $            &  $ -0.30 \pm 0.11 $  &  SNR? \\
    G28.22$-$0.09  &  $ < 1.5 $         &  $ 0.04 \pm 0.01 $  &  $ 0.05 \pm 0.01 $  &  $ < 10 $         &  $ +0.23 \pm 0.38 $  &  \\
    G28.33$+$0.06  &  $ - $             &  $ 0.56 \pm 0.11 $  &  $ 0.42 \pm 0.06 $  &  $ - $            &  $ -0.18 \pm 0.18 $  &  \\
    G28.36$+$0.21  &  $ 3.6 \pm 0.5 $   &  $ 2.2 \pm 0.4 $    &  $ 1.4 \pm 0.4 $    &  $ 16 \pm 18 $    &  $ -0.28 \pm 0.11 $  &  SNR: \S \ref{subsec:G28.36} \\   
    G28.56$+$0.00  &  $ 0.64 \pm 0.15 $ &  $ 0.89 \pm 0.10 $  &  $ 0.77 \pm 0.09 $  &  $ 5.2 \pm 7.1 $  &  $ +0.07 \pm 0.08 $  &  \\
    G28.64$+$0.20  &  $ - $             &  $ 2.4 \pm 0.3 $    &  $ 2.2 \pm 0.2 $    &  $ < 18 $         &  $ -0.14 \pm 0.09 $  &  (1,2) \\
    G28.78$-$0.44  &  $ 2.2 \pm 0.2 $   &  $ 1.1 \pm 0.1 $    &  $ 0.55 \pm 0.07 $  &  $ 4.4 \pm 1.9 $  &  $ -0.42 \pm 0.04 $  &  SNR: \S \ref{subsec:G28.78}, (1) \\   
    G28.88$+$0.41  &  $ 1.8 \pm 0.5 $   &  $ 2.2 \pm 0.6 $    &  $ 2.1 \pm 0.8 $    &  $ 3.3 \pm 6.0 $  &  $ +0.02 \pm 0.17 $  &  \\
    G28.92$+$0.26  &  $ - $             &  $ - $              &  $ - $              &  $ - $            &  $ - $               &  (3) \\
    G29.38$+$0.10  &  $ 2.6 \pm 0.3 $   &  $ 2.5 \pm 0.2 $    &  $ 1.6 \pm 0.2 $    &  $ 5.5 \pm 1.0 $  &  $ -0.04 \pm 0.07^\dagger $  &  SNR: \S \ref{subsec:G29.38}, (1) \\   
                   &                    &                     &                     &                   &  $ -0.35 \pm 0.07^\dagger $  &  \\
    G29.41$-$0.18  &  $ - $             &  $ 1.27 \pm 0.37 $  &  $ 0.94 \pm 0.24 $  &  $ - $            &  $ -0.21 \pm 0.29 $  &  \\
    G29.92$+$0.21  &  $ < 1.4 $         &  $ 0.28 \pm 0.05 $  &  $ < 0.12 $         &  $ < 1.7 $        &  $ \sim -0.5 ~\mathrm{to} -0.8 $  &  SNR? \\
    G31.22$-$0.02  &  $ - $             &  $ - $              &  $ - $              &  $ - $            &  $ - $               &  (4) \\
    G31.44$+$0.36  &  $ - $             &  $ - $              &  $ - $              &  $ - $            &  $ - $               &  (5) \\
    G31.93$+$0.16  &  $ 0.24 \pm 0.02 $ &  $ 0.10 \pm 0.04 $  &  $ 0.05 \pm 0.01 $  &  $ < 80 $         &  $ -0.45 \pm 0.07 $  &  SNR? \\
    G32.22$-$0.21  &  $ 0.36 \pm 0.04 $ &  $ 0.54 \pm 0.09 $  &  $ 0.48 \pm 0.05 $  &  $ - $            &  $ +0.10 \pm 0.08 $  &  \\
    G32.37$-$0.51  &  $ < 14 $          &  $ < 33 $           &  $ < 4.4 $          &  $ < 16 $         &  $ - $               &  \\
    G32.73$+$0.15  &  $ 0.41 \pm 0.08 $ &  $ 0.13 \pm 0.07 $  &  $ - $              &  $ - $            &  $ -0.68 \pm 0.42 $  &  SNR? \\
    G33.62$-$0.23  &  $ 0.19 \pm 0.05 $ &  $ 0.15 \pm 0.05 $  &  $ < 0.22 $         &  $ < 27 $         &  $ -0.15 \pm 0.46 $  &  \\
    G33.85$+$0.06  &                    &                     &                     &                   &                      &  point sources \\
    G34.93$-$0.24  &  $ 0.73 \pm 0.17 $ &  $ 0.64 \pm 0.26 $  &  $ 0.82 \pm 0.35 $  &  $ - $            &  $ -0.00 \pm 0.19 $  &  \\
\hline
\end{tabular}
\tablefoot{$S_{\mathrm{0.2GHz}}$, $S_{\mathrm{1.4GHz}}$, $S_{\mathrm{5.8GHz}}$, and $p_{\mathrm{5.8GHz}}$ are as defined in Table~\ref{tab:knownSNRs}.  \\
($^\dagger$)~ Since a break in the spectrum is clearly visible, we report both the spectral indices.\\  
(1) The spectral index of G28.64+0.20, G28.78$-$0.44 and G29.38+0.10 was derived using flux densities from the Effelsberg 11~cm and the Nobeyama 3~cm surveys as well.\\  
(2) The flux densities of G28.64$+$0.20 are measured only for the arc-shaped structure on the West.\\
(3) G28.92$+$0.26 is resolved to the GLOSTAR candidates G028.929+0.254 and G028.877+0.241.\\  
(4) We study G31.22$-$0.02 as the GLOSTAR candidate G031.256$-$0.041.\\
(5) No measurement possible in any survey, either due to artefact contamination or insufficient sensitivity.
}
\end{table*}

\begin{table*}
\centering
\caption{Flux densities and spectral indices of the GLOSTAR SNR candidates in the pilot region.}
\label{tab:GLOcands}
\begin{tabular}{l c c c c c l}
\hline\hline
Name & $S_{\mathrm{0.2GHz}}$ & $S_{\mathrm{1.4GHz}}$ & $S_{\mathrm{5.8GHz}}$ & $p_{\mathrm{5.8GHz}}$ & $\alpha_\mathrm{FD}$ & Remarks \\
& (Jy) & (Jy) & (Jy) & (\%) & & \\
\hline
    G028.524$+$0.268  &  $ 0.50 \pm 0.08 $  &  $ 0.54 \pm 0.06 $   &  $ 0.54 \pm 0.07 $   &  $ < 18 $     &  $ +0.03 \pm 0.06 $  &  \\
    G028.870$+$0.616  &  $ 0.10 \pm 0.01 $  &  $ 0.019 \pm 0.017 $ &  $ 0.026 \pm 0.011 $ &  $ < 100 $    &  $ -0.43 \pm 0.31 $  &  \\
    G028.877$+$0.241  &  $ - $              &  $ 0.023 \pm 0.016 $ &  $ 0.015 \pm 0.011 $ &  $ < 47 $     &  $ -0.33 \pm 0.87 $  &  \\
    G028.929$+$0.254  &  $ 0.74 \pm 0.17 $  &  $ 0.28 \pm 0.03 $   &  $ 0.27 \pm 0.03 $   &  $ < 17 $     &  $ -0.31 \pm 0.14 $  &  \\
    G029.329$+$0.280  &  $ < 0.64 $         &  $ < 0.26 $          &  $ 0.08 \pm 0.02 $   &  $ < 49 $     &  $ > -0.63 $         &  \\
    G030.303$+$0.128  &  $ - $              &  $ < 0.04 $          &  $ 0.021 \pm 0.003 $ &  $ < 33 $     &  $ > -0.38 $         &  \\
    G030.362$+$0.623  &  $ < 0.8 $          &  $ < 1.2  $          &  $ 0.11 \pm 0.02 $   &  $ < 100 $    &  $ > -0.59 $         &  \\
    G030.375$+$0.424  &  $ - $              &  $ 0.04 \pm 0.01 $   &  $ 0.02 \pm 0.01 $   &  $ < 17 $     &  $ -0.58 \pm 0.54 $  &  \\
    G030.508$+$0.574  &  $ - $              &  $ - $               &  $ 0.07 \pm 0.03 $   &  $ < 100 $    &  $ - $               &  \\
    G031.256$-$0.041  &  $ \sim 0.4 $       &  $ 0.33 \pm 0.07 $   &  $ 0.37 \pm 0.05 $   &  $ - $        &  $ +0.04 \pm 0.10 $  &  PWN?: \S \ref{subsec:G031.256}\\
    G032.458$-$0.112  &  $ - $              &  $ 0.03 \pm 0.01 $   &  $ 0.03 \pm 0.01 $   &  $ < 63 $     &  $ +0.11 \pm 0.52 $  &  \\
    G034.524$-$0.761  &  $ 0.83 \pm 0.14 $  &  $ 0.18 \pm 0.03 $   &  $ 0.04 \pm 0.01 $   & $ 13 \pm 10 $ &  $ -0.93 \pm 0.15 $  &  SNR?: \S \ref{subsec:G034.524} \\
    G034.619$+$0.240  &  $ 0.30 \pm 0.10 $  &  $ 0.20 \pm 0.03 $   &  $ 0.20 \pm 0.02 $   &  $ < 31 $     &  $ -0.09 \pm 0.16 $  &  \\
    G035.129$-$0.343  &  $ < 0.11 $         &  $ 0.04 \pm 0.01 $   &  $ 0.033 \pm 0.005 $ &  $ < 82 $     &  $ -0.17 \pm 0.16 $  &  \\
\hline
\end{tabular}
\tablefoot{$S_{\mathrm{0.2GHz}}$, $S_{\mathrm{1.4GHz}}$, $S_{\mathrm{5.8GHz}}$, and $p_{\mathrm{5.8GHz}}$ are as defined in Table~\ref{tab:knownSNRs}.}
\end{table*}

\begin{figure*}
    \centering
    \includegraphics[width=0.9\textwidth]{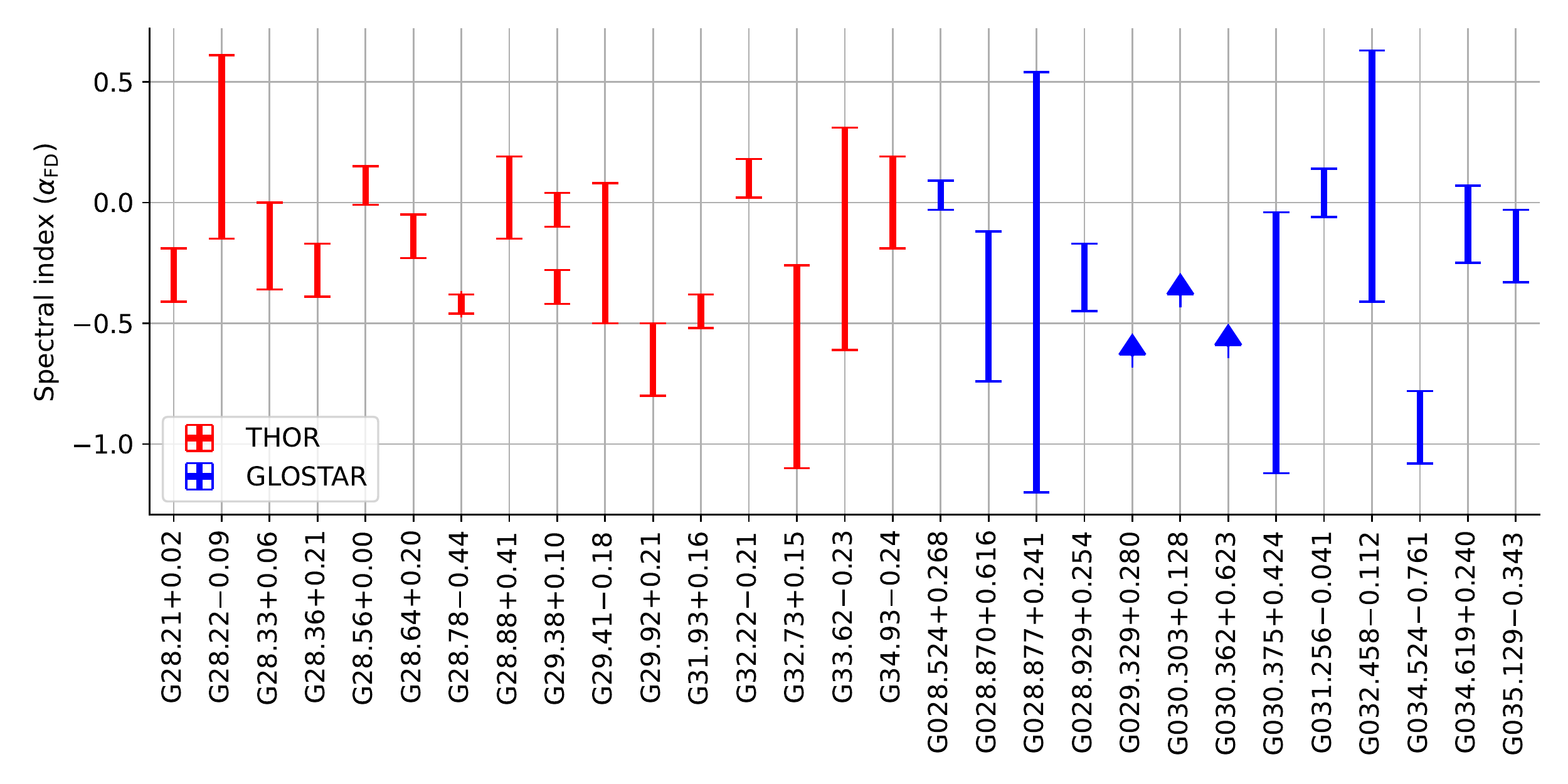}
    \caption{Flux density spectral indices ($\alpha_{\mathrm{FD}}$) of the candidate SNRs being studied in this work.  Candidates with lower limits are represented by upward arrows.  Since G29.38+0.10 has a spectral break, both the spectral indices are shown. }
    \label{fig:spidxdist}
\end{figure*}

\begin{figure*}
    \centering
    \includegraphics[width=16cm]{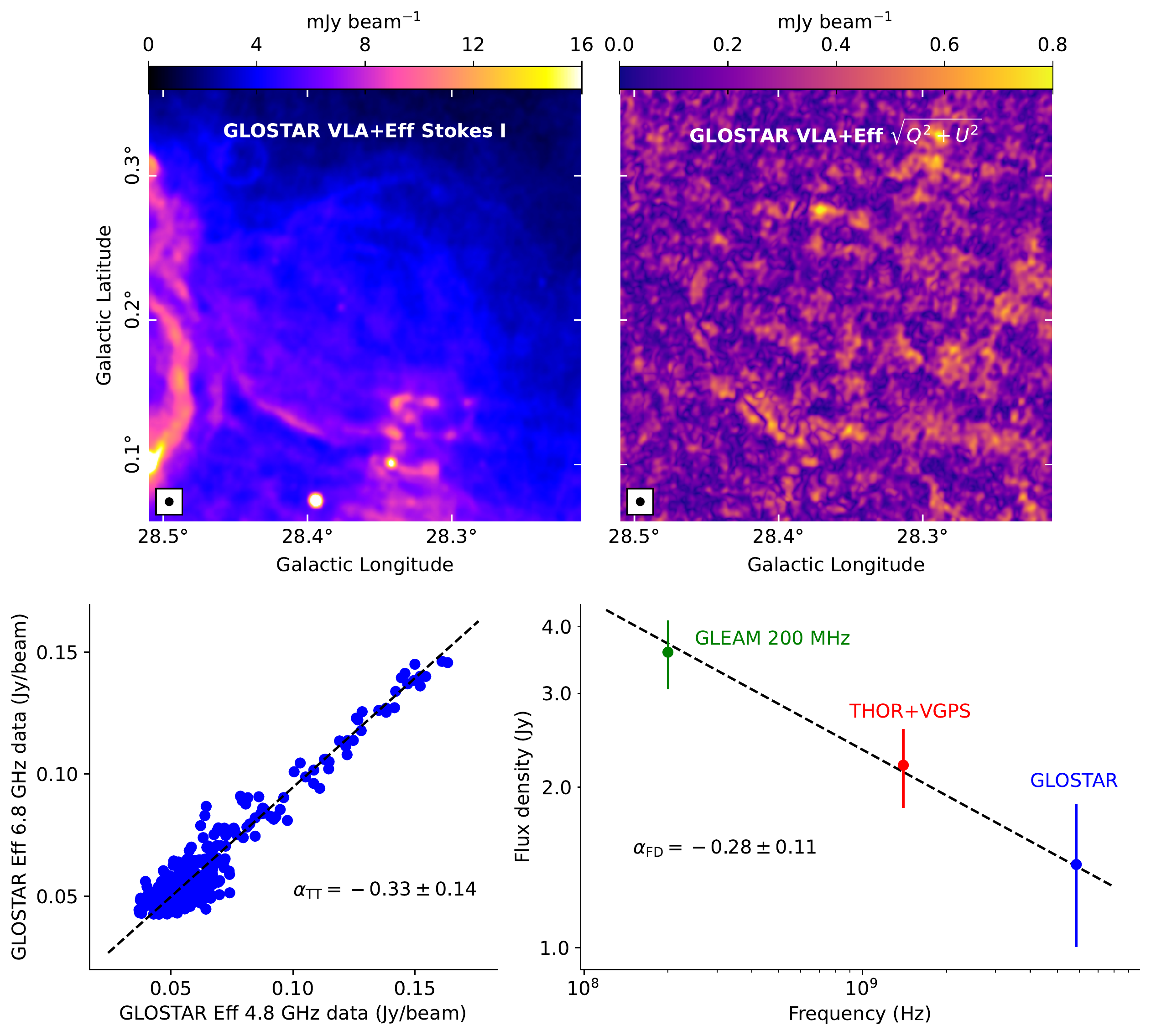}
    \caption{ Same as Fig.~\ref{fig:G35.6-0.4} but for SNR G28.36+0.21. }
    \label{fig:G28.36}
\end{figure*}

\begin{figure*}
    \centering
    \includegraphics[width=16cm]{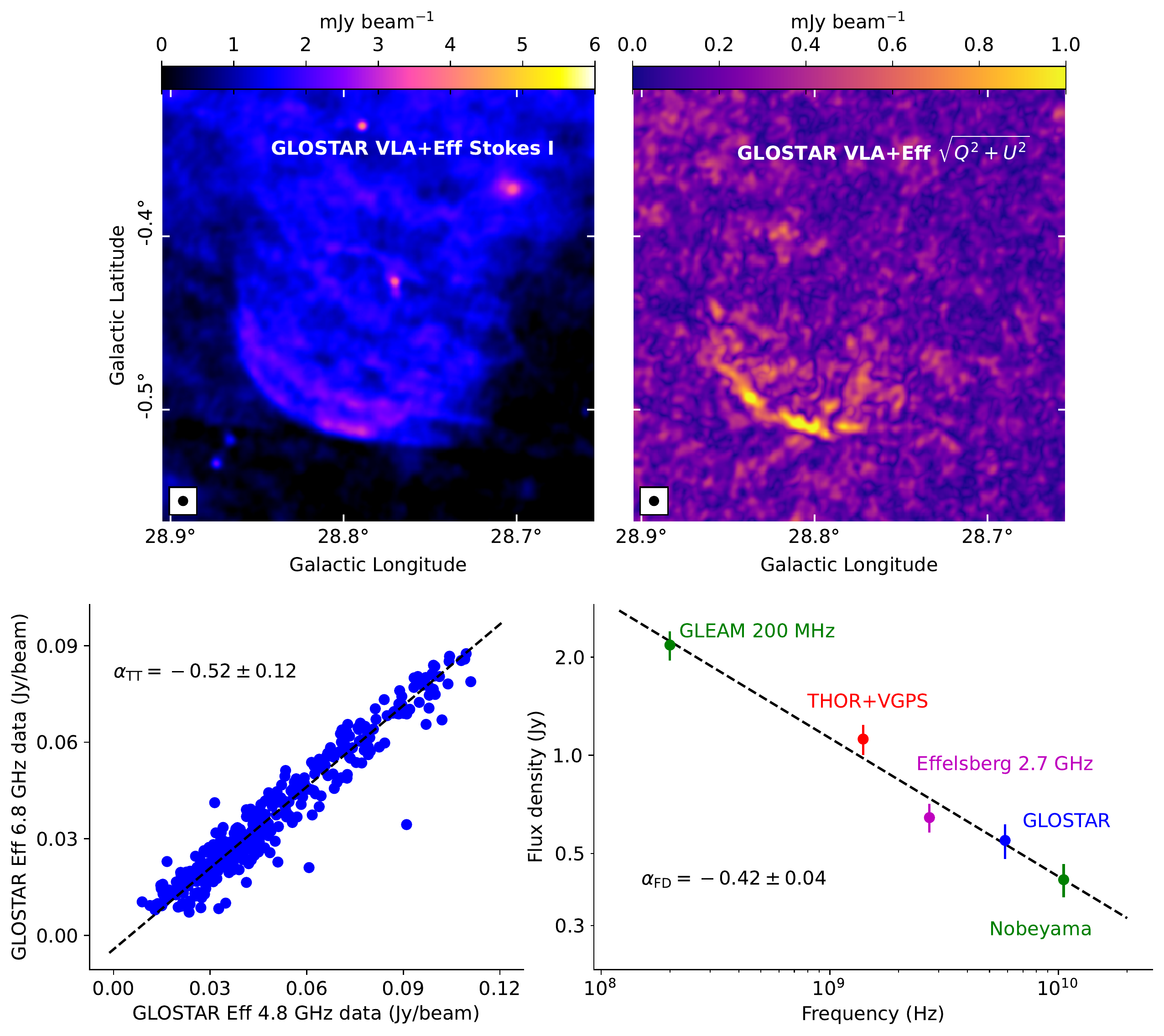}
    \caption{Same as Fig.~\ref{fig:G35.6-0.4} but for SNR G28.78-0.44.}
    \label{fig:G28.78}
\end{figure*}

\begin{figure*}
    \centering
    \includegraphics[width=16cm]{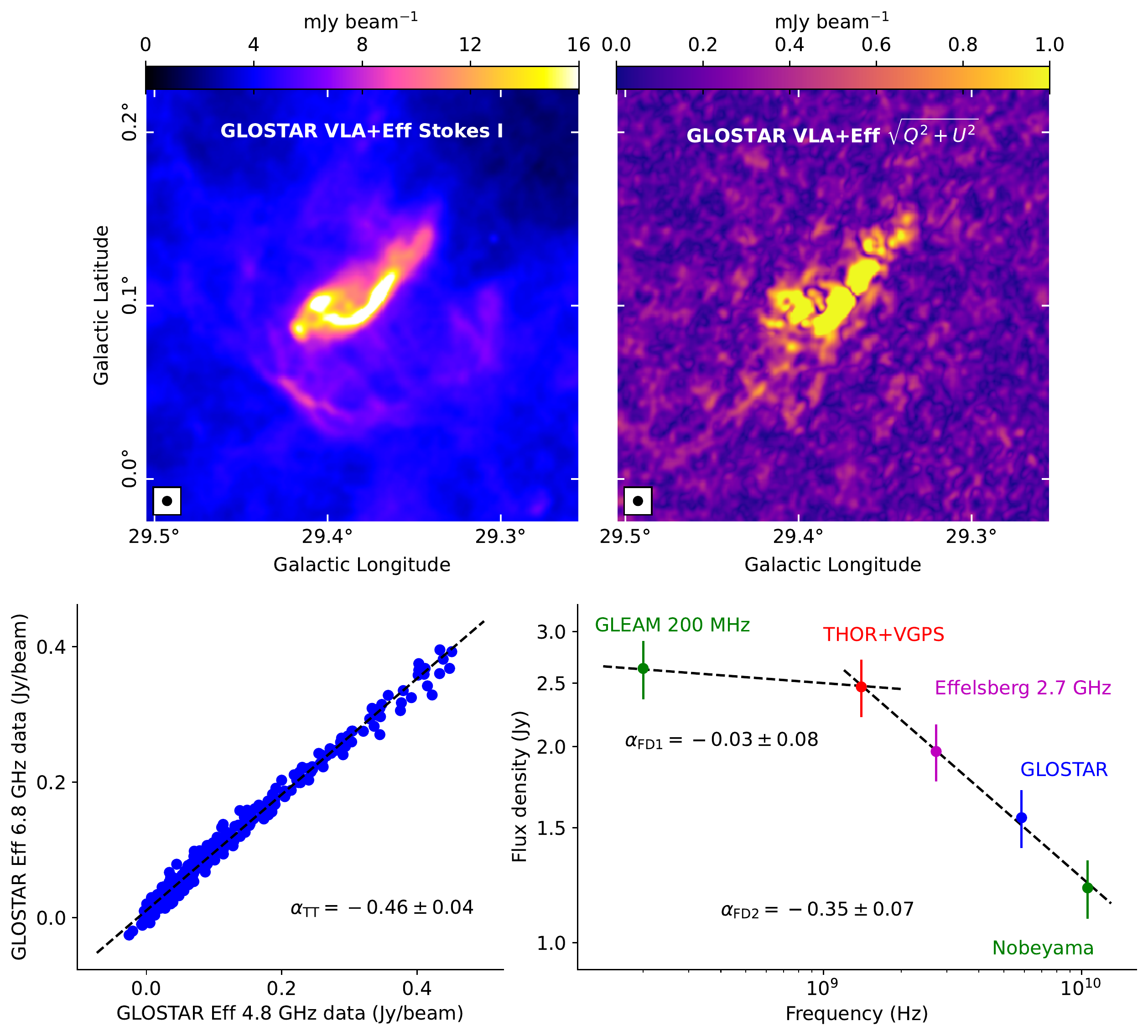}
    \caption{Same as Fig.~\ref{fig:G35.6-0.4} but for SNR G29.38+0.10.}
    \label{fig:G29.38}
\end{figure*}

In the pilot region, we had discovered 14 new candidate SNRs from the GLOSTAR survey in our previous work (\dok), in addition to the 21 candidates discovered by \loren ~using THOR+VGPS images.  The continuum flux densities of these candidates are presented in Table~\ref{tab:THORcands} (from THOR+VGPS) and Table~\ref{tab:GLOcands} (from GLOSTAR).  We derived flux density spectral indices whenever possible, and these are plotted in Fig.~\ref{fig:spidxdist}.  We discuss five objects for which there is good evidence of nonthermal emission in detail in the following sections.  We also find that 14 other candidates possibly have a negative spectral index.  But since they are quite faint and the morphology of these candidates is not clear (see Figs.~\ref{fig:G28.21} and \ref{fig:G31.93}), we do not discuss them further.

\subsection{G28.36+0.21}
\label{subsec:G28.36}

First identified by \loren ~as a SNR candidate using the THOR+VGPS images, G28.36+0.21 has a limb-brightened structure that is typical of SNRs.  \citet{2019PASA...36...47H} noted this object as a high confidence level candidate, deriving a spectral index of ${\sim}-0.7$ from $70-230$~MHz.  We detect this object in the Stokes~$I$ images of our GLOSTAR survey (Fig.~\ref{fig:G28.36}), with the same morphology as observed in the THOR+VGPS images.  Its fractional polarization is about $2\%$, which is not unusual in SNRs \citep[e.g.,][]{2011A&A...536A..83S}.  The linearly polarized intensity map from GLOSTAR shows a faint structure, close to the noise level in this region, that resembles the total intensity of the shell of this object.  From the Effelsberg images of our survey, we made a TT-plot and obtained a spectral index of $-0.33 \pm 0.14$.  By measuring the background-subtracted flux densities in the images of GLOSTAR combination, THOR+VGPS and GLEAM, we obtain a brightness spectral index of $-0.28 \pm 0.11$.  These measurements and the morphology we observe in the total and linearly polarized intensity images provide ample evidence of nonthermal emission from this object, and hence we conclude that G28.36+0.21 is indeed a SNR.

\subsection{G28.78-0.44}
\label{subsec:G28.78}

The candidate SNR G28.78-0.44 (Fig.~\ref{fig:G28.78}) had previously been identified in the MAGPIS and the THOR+VGPS surveys (\citealt{2006AJ....131.2525H}; \loren).  \citet{2019PASA...36...47H} derive a spectral index of ${\sim}-0.7$ in their GLEAM survey (70--230~MHz), consistent with the spectral index from the TIFR-GMRT Sky Survey and the NRAO VLA Sky Survey \citep{2018MNRAS.474.5008D,2018ApJ...866...61D}.  While the polarization from this object was already clearly visible in the VLA images of the GLOSTAR survey (\dok), the addition of the Effelsberg data allows us to measure its flux densities at 5.8~GHz.  The fractional polarization we measure in the combination images is about $4\%$.  We also detect this object in the Effelsberg 11~cm survey \citep{1984A&AS...58..197R} and the Nobeyama 10~GHz survey \citep{1987PASJ...39..709H}.  These give us a broadband flux density spectral index of $-0.42 \pm 0.04$, which is consistent with the TT-plot spectral index from the Effelsberg images of the GLOSTAR survey alone ($-0.52 \pm 0.12$, see Fig.~\ref{fig:G28.78}).  Thus we find strong evidence that this filled-shell object is a SNR.

\subsection{G29.38+0.10}
\label{subsec:G29.38}

This source appears to have a complex structure with a bright pulsar wind nebula (PWN) and a faint SNR shell in the GLOSTAR combination image (Fig.~\ref{fig:G29.38}).  The central structure of this complex is bright and highly polarized in the combination images, with the degree of linear polarization reaching as high as $30\%$ in some pixels.  For the whole complex, this value is $5.5 \pm 0.8\%$.  We had detected strong linear polarization from this object in our previous work as well (\dok), which was based only on the D-configuration VLA images.  Its low frequency spectral index measured using the GLEAM images by \citet{2019PASA...36...47H} for the whole complex, and for the central source by \citet{2018ApJ...866...61D} using the TGSS-NVSS spectral index map \citep{2018MNRAS.474.5008D} is approximately zero, which is typical of PWNe.  We calculate a similar spectral index using the THOR+VGPS and GLEAM images as well.  However, between the THOR+VGPS and the GLOSTAR combination images, the flux density falls with a spectral index of $\alpha_{\mathrm{FD}}{\sim}-0.34$.  Constructing a TT-plot using images from the two bands of the GLOSTAR Effelsberg data, we measure a value $\alpha_{\mathrm{TT}}{\sim}-0.46$.  This implies that there is a break in the spectrum of this source near 2~GHz, or a gradual turnover.  Such a varying spectral index at these frequencies is once again typical of PWNe \citep[see][]{1973ApJ...186..249P,1984ApJ...278..630R,2011A&A...536A..83S}.  These facts provide further evidence that G29.38+0.10 is a PWN+SNR shell complex.

\subsection{G031.256-0.041}
\label{subsec:G031.256}

\begin{figure*}
    \centering
    \includegraphics[width=\textwidth]{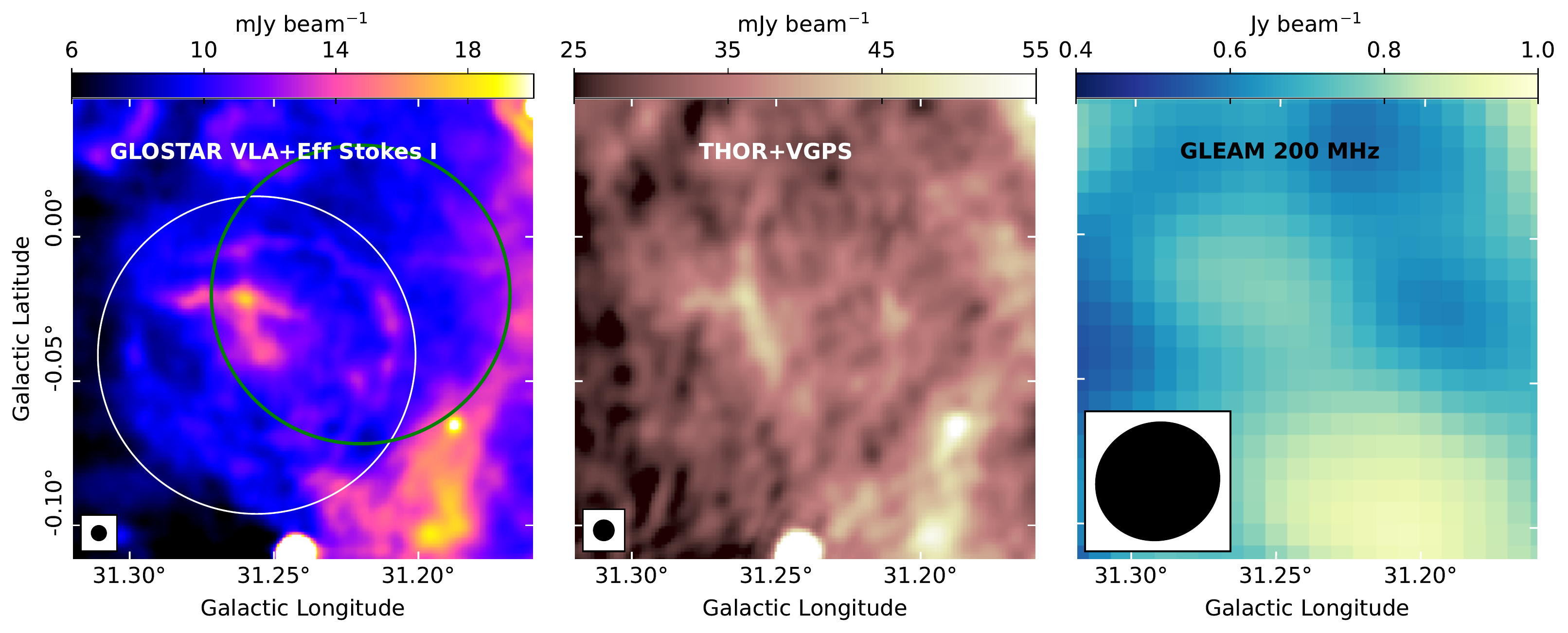}
    \caption{GLOSTAR SNR candidate G031.256-0.041 as seen in the GLOSTAR combination, THOR+VGPS, and the 200~MHz GLEAM images.  The white circle marks the extent of the GLOSTAR candidate G031.256-0.041, whereas the green circle marks the THOR SNR candidate G31.22-0.02. }
    \label{fig:G031.256}
\end{figure*}

\begin{figure*}
    \centering
    \includegraphics[width=16cm]{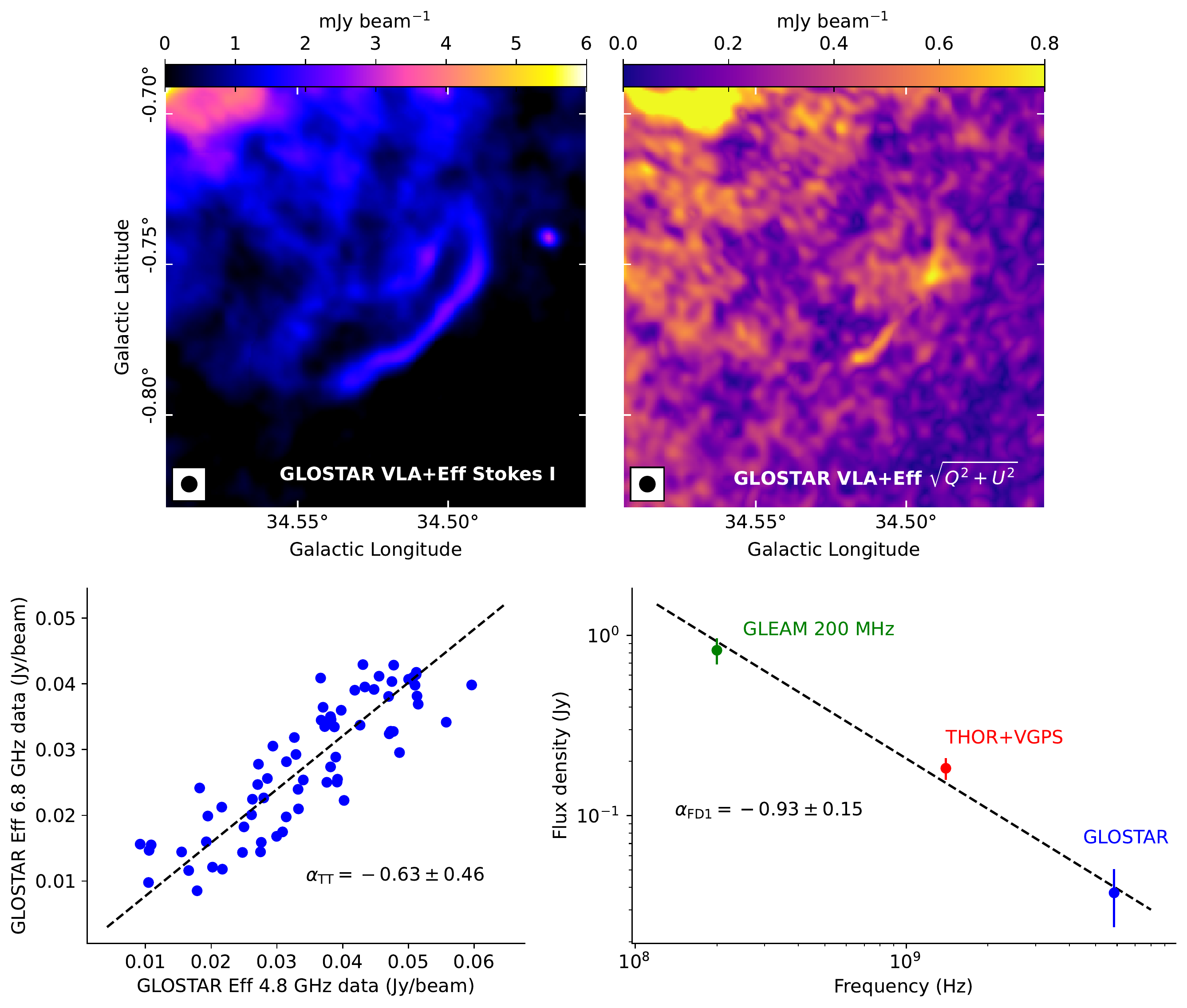}
    \caption{Same as Fig.~\ref{fig:G35.6-0.4} but for GLOSTAR SNR candidate G034.524-0.761.}
    \label{fig:G034.524}
\end{figure*}

\loren ~cataloged G31.22-0.02 as a shell-shaped SNR candidate based on the THOR+VGPS images.  It lies in a crowded field with a strong background, due to which the determination of the TT-plot spectral index from the Effelsberg images ($\alpha_\mathrm{TT}$) was not possible.  This region is better resolved in the GLOSTAR combination images, in which we identify the brightest part of the supposed shell of G31.22-0.02 (at $l{\sim}31.26\degree$, $b{\sim}-0.02\degree$) to be inside another shell (Fig.~\ref{fig:G031.256}).  We believe that this is a PWN+shell complex, and named it as a GLOSTAR SNR candidate G031.256-0.041 in our previous work (\dok).  The flux densities we measured in the THOR+VGPS and the GLOSTAR combination images are similar within uncertainties ($S{\sim}0.35$\,Jy), giving a spectral index close to zero between 1.4~GHz and 5.8~GHz.  In the 200~MHz GLEAM images, what we believe is the center of the PWN (at $l{\sim}31.26\degree$, $b{\sim}-0.02\degree$) is barely resolved, with a peak brightness of nearly 0.8~Jy\,beam$^{-1}$.  The background level in this region is about 0.4~Jy, implying that the flux density of the peak is ${\sim}0.4$~Jy, similar to the flux densities from the GLOSTAR combination and the THOR+VGPS images.  Unfortunately, the linearly polarized intensity images from GLOSTAR in this region are contaminated with sidelobe artefacts of nearby bright sources, prohibiting us from measuring its degree of polarization.  The morphology and the estimated spectral index are, however, consistent with our PWN+SNR shell interpretation.

\subsection{G034.524-0.761}
\label{subsec:G034.524}

We discovered the SNR candidate G034.524-0.761 in our previous GLOSTAR work, where we had identified clear linear polarization from the VLA data (see Fig.~11 of \dok).  With the addition of the Effelsberg data to the VLA images, we now obtain a degree of polarization ${\sim}10\%$ from this candidate.  In addition, we obtain a TT-plot spectral index of ${\sim}-0.6$ using the Effelsberg images, although with a large uncertainty of ${\sim}0.5$.  We measured flux densities in the 200~MHz GLEAM and the 1.4~GHz THOR+VGPS images, which give us a spectral index of ${\sim}-0.9$.  While all these facts point to a nonthermal origin of the emission from this region, the morphology of this candidate (Fig.~\ref{fig:G034.524}) indicates that this might be a filament.  For this reason, we cannot conclude that this object is a SNR.

\section{Discussion}
\label{sec:disc}

It is evident from Fig.~\ref{fig:spidxdist} that the spectral indices of several SNR candidates are not well constrained yet.  Most of them have a small angular size and a low surface brightness, and they lie in crowded regions with a strong background; these conditions result in large uncertainties in the measurement of their spectral indices.  Moreover, the polarization signals from several SNRs may remain undetected because of limited sensitivity \citep[the linearly polarized flux density is typically only a few percent of the total flux density, e.g.,][]{2011A&A...536A..83S}.  Deeper observations of these candidates across the radio band are necessary to constrain their spectral indices and linear polarization better.  However the current results do not look very promising since the rate of confirmation appears to be quite low, and we are forced to ponder over the strategy to identify new SNRs.

Since most of the bright SNRs are likely to have been discovered already, it might progressively get more difficult to find the remaining fainter ones.  H~{\footnotesize{II}} regions are more numerous in the Galaxy, and there is a chance that the fainter H~{\footnotesize{II}} regions contaminate the sample of the faint SNRs.  However, the SNR candidates identified by \loren ~and \dok ~do not have any significant coincident MIR emissions detected in the {\textit{Spitzer}} MIR surveys, which can detect H~{\footnotesize{II}} regions anywhere in the Galaxy \citep{2014ApJS..212....1A}.  Hence, we believe that, if the SNR candidates do not turn out to be SNRs, the confusion must be due to radio emitters other than H~{\footnotesize{II}} regions, although it is unclear what kind of objects they might be.  \loren ~and \dok ~suggest that the remaining undetected SNRs must be faint and also have a small angular size.  We turn our attention toward these properties of the sample of the SNR candidates.

\subsection{Angular radius}

One question that needs to be answered before starting the search for the remaining SNRs is whether most of them are indeed small, since that would determine what resolution is necessary to detect the `missing' SNRs.  To estimate their apparent angular extents, we ran a simple Monte Carlo simulation of evolution of SNRs in the Milky Way.  SNRs are evolved in a locally uniform ISM using the expressions from \citet{2011piim.book.....D}, which are based on the four classical stages as proposed by \citet{1972ARA&A..10..129W}:  
\begin{enumerate}
    \item The earliest part of the evolution is known as the free-expansion or the ejecta-dominated phase.  We assume that the mass of the swept up ISM ($m_\mathrm{sw}$) is negligible compared to the mass of the SN ejecta ($m_\mathrm{ej}$) in this stage.
    \item Sedov-Taylor phase begins when the shocked and swept up mass is comparable to the ejecta mass $m_\mathrm{sw} \sim m_\mathrm{ej}$, during which the explosion can be approximated as a point source injecting only energy.
    \item Snowplow phase begins when the radiative cooling losses become important and the matter behind the SNR shock cools rapidly to form a cold and dense shell.  In the hot and tenuous medium that is interior to the shock, however, the energy losses do not yet play a role, and the pressure from this hot central volume drives the momentum of the dense outer shell.
    \item The final phase is `dispersion' as the SNR merges into the surrounding ISM and fades away when the shock speed drops to the ambient velocity dispersion levels.
\end{enumerate}

We derive the radius of each SNR based on the time since explosion and the position in the Galaxy.  Following are the main parameters and inputs of the simulation:
\begin{itemize}
    \item Galactic supernova rate of one per 40 years, with the core-collapse and thermo-nuclear types being $85\%$ and $15\%$ respectively \citep{1994ApJS...92..487T,2005AJ....130.1652R}.
    \item Three dimensional gas density model of the Milky Way from \citet{2006A&A...459..113M}.
    \item A random Monte Carlo model of the two-dimensional distribution of supernova events in a disk with a central hole and a two-arm spiral following \citet{1991ApJ...378...93L}.  The central hole is to account for the dearth of massive star formation, and by extension SNRs, near the Galactic center \citep[see][for example]{2021A&A...651A..88N,2022arXiv220904570R}.
    \item Core-collapse SN events, which trace massive star formation, are chosen to have a scale height of 80~pc, the same as the scale height of the molecular gas \citep[from][]{2006A&A...459..113M}.  
    \item Type~Ia SNe arise due to mass accretion onto old degenerate stars; accordingly we use the thick disk scale height of 0.7~kpc from \citet{2011A&A...535A.107K}.
    \item The maximum lifetime of SNRs is fixed at 80\,000 years \citep{1994ApJ...437..781F}.
    \item For a Type~Ia supernova, the kinetic energy of the ejecta is fixed at $10^{51}$~erg and the ejecta mass is normally distributed from 0.8$M_\odot$--1.8$M_\odot$ \citep[following][]{2014MNRAS.445.2535S}.
    \item For the more numerous core-collapse supernova events, the ejecta mass (8$M_\odot$--11$M_\odot$) and the kinetic energy (0.2--1.3 times $10^{51}$~erg) are randomly drawn from distributions adapted from the results of \citet{2022A&A...660A..41M}.
\end{itemize}

There are, however, some caveats to consider:
\begin{itemize}
    \item Realistically, the properties of the ISM are not smoothly varying functions of position as the model given by \citet{2006A&A...459..113M}.  The ISM number density can drastically change depending on the environment, especially in the case of previous mass-loss events such as stellar winds.  These affect the evolution of SNRs in a crucial and nontrivial manner \citep[e.g.,][]{2021ApJ...919L..16Y}. 
    \item The distribution of supernova events follows the model of \citet{1991ApJ...378...93L}, which is quite simplistic.  But similar to their findings, we also observe that the results are insensitive to parameters of the disk and the spiral arms.  The inverse dependence of angular radius with distance makes our result even more robust than that of \citet{1991ApJ...378...93L}.
    \item The distributions of ejecta mass we used \citep[from][]{2014MNRAS.445.2535S,2022A&A...660A..41M} may not hold for the Milky Way accurately, since those results are from the nearby local universe with supernovae from several galaxies.  However, we find that even if the ejecta mass for core-collapse supernovae was only 1$M_\odot$ instead of 8$M_\odot$--11$M_\odot$, the results are mostly the same.
    \item There is evidence that the explosion energies of supernovae can have a range wider than that we have taken, for both Type~Ia and core-collapse, from ${\sim}10^{49}$ to ${\sim}10^{52}$~erg \citep[e.g.,][]{2005ApJ...623.1011B,2015ApJ...805..150F,2015ApJ...801...90P,2019MNRAS.489..641M,2020ApJS..248...16L}.  Even with a wider range, we find that the resultant radius distribution does not significantly change. 
    \item We do not take into account the effects of clustering.  This is the main drawback of this simulation.  A significant fraction of massive star formation---and the number of SN events by extension---happens in clusters \citep[e.g.,][]{2014PhR...539...49K}.  \citet{2001RvMP...73.1031F} estimates that ${\sim}60$\% of O stars probably remain in their natal group, while the rest of them end up in the `field'.  If multiple supernovae occur in succession in such clusters, this might result in the formation of a super-bubble \citep[e.g.,][]{2013A&A...550A..23E}.
\end{itemize}

\begin{figure}
    \centering
    \includegraphics[width=7cm]{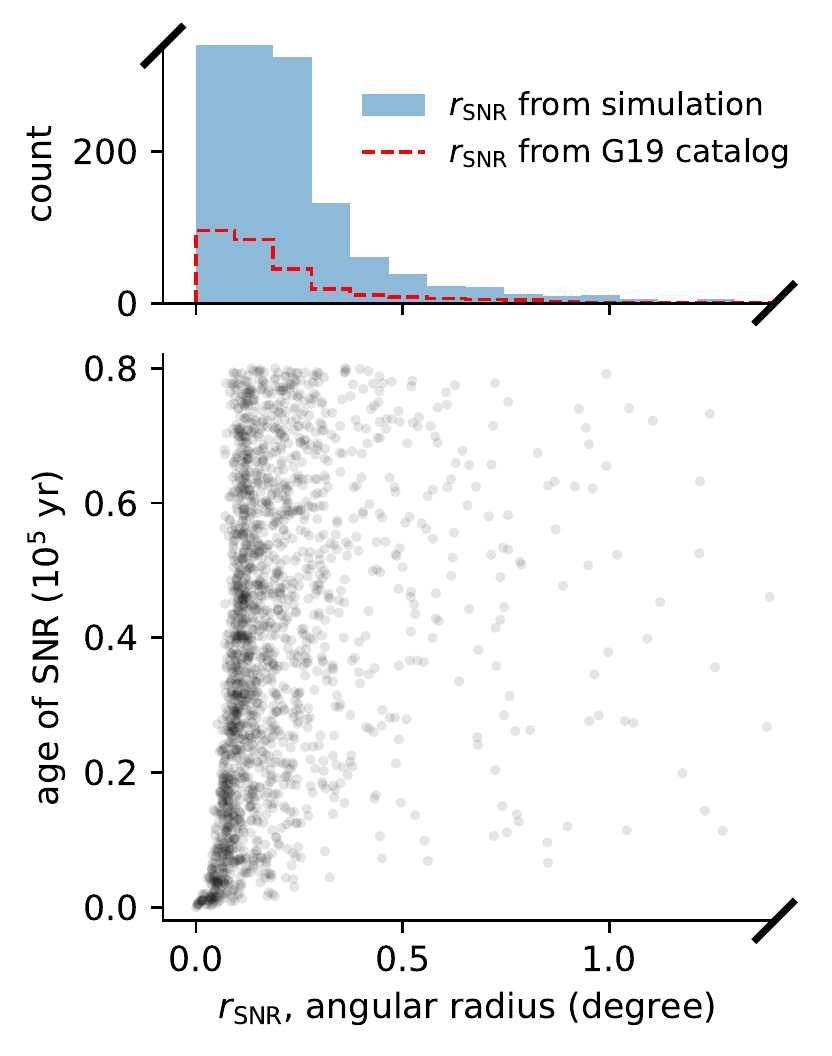}
    \caption{ Snapshot of the age and the angular radius distribution of SNRs (as seen from the Sun) from the simulation, at a time of 1.8 million years, is shown in the bottom panel.  The angular radius distribution is shown on the top panel in blue, along with the distribution of the SNRs in the catalog of \citet{2019JApA...40...36G} in red dashed lines.  The axes are clipped to show the distributions better. }
    \label{fig:simsnap}
\end{figure}

We ran the simulation for two million years, which is several generations of SNRs.  A snapshot at a time of 1.8~million years is presented in Fig.~\ref{fig:simsnap}, and a movie of the whole two million years is \href{https://cloud.mpifr-bonn.mpg.de/index.php/s/m3TxJESGs7tNQxo}{available online}.  Given that the lifetime of a SNR and the SN rate are fixed at 80,000 years and one for every 40 years respectively, about 2000 SNRs exist at the end of the simulation.  It is clear that most of the SNRs are quite small with angular radii of only a few arcminutes, similar to the THOR and GLOSTAR SNR candidates.  Even if the lifetime of a typical SNR is longer than 80,000 years as we had used, the resultant distribution does not shift to higher angular scales significantly.  This is due to the fact that the expansion is considerably slower in the later stages of SNR evolution.  While this simulation only serves as a first approximation since we do not consider several effects such as those mentioned above, it is nevertheless useful to give us an idea of what to expect.  And the result reiterates the views of \loren ~and \dok ~that SNR searches must focus on small angular sized objects to make the most gains.

\subsection{Radio surface brightness}

In the simulation described above, we also measured the area of overlap of SNRs.  We find it to be typically less than 10\% of the total sky area covered by SNRs, suggesting that the confusion due to SNRs overlapping themselves may not be important.  However, the SNRs originating from core-collapse events are located near massive star forming complexes, which also contain other extended structures emitting at radio wavelengths.  H~{\footnotesize{II}} regions are the most likely sources of positional overlapping confusion: they are probably over 8000 in number \citep{2014ApJS..212....1A}, and the range of the values their radio surface brightness is similar to that of SNRs. 

Currently, the faintest SNR known has a brightness temperature of about 0.33~K at 1~GHz \citep{2017A&A...597A.116K}, and, by extrapolating to 1~GHz assuming a nonthermal spectral index, we find that the SNR candidates from \loren ~and \dok ~are at a similar or lower surface brightness.  On the other hand, the background emission from the diffuse gas in the Milky Way is at a level of a few Kelvin in the inner Galactic plane at 1~GHz \citep[e.g.,][]{1990A&AS...83..539R}, and it is even higher in regions such as the mini-starburst W43 where one expects many SNRs due to recent massive star formation activity.  This implies that the diffuse background emission is a critical source of confusion, and finding new SNRs will probably be more difficult from now on.  Interferometric surveys at lower frequencies, such as MeerKAT, appear promising in the search for new SNRs \citep[e.g.,][]{2022ApJ...925..165H}, but the nonthermal Galactic background emission is also stronger at lower frequencies and may contribute to the confusion.

\section{Summary and Conclusions}
\label{sec:conclusions}

We derived spectral indices of previously confirmed SNRs in the Galactic longitude range $28\degree<l<36\degree$, using the VLA-D+Effelsberg combination images of the 4--8~GHz GLOSTAR survey in addition to other complementary and archival survey data.  These include the first radio spectral index determinations for SNRs G32.1-0.9 and G32.4+0.1, along with the first reported spectral break for SNR G35.6-0.4.  We showed that G31.5-0.6 may not be a SNR, and we provided further evidence of nonthermal emission from the SNR candidates G28.36+0.21, G28.78-0.44, G29.38+0.10, and G034.524-0.761.  We find that G28.36+0.21 and G28.78-0.44 are typical SNR shells, and G29.38+0.10 is a PWN+shell complex.  Based on a simple Monte-Carlo simulation of SN events in the Milky Way, we find that most of the SNRs yet to be discovered must have angular sizes smaller than half a degree.  Hence, despite the low rate of confirmation, we believe that future studies must focus on small angular sized objects such as the THOR and GLOSTAR SNRs.  The forthcoming Effelsberg images from the GLOSTAR survey for the rest of the coverage will be analyzed in the coming months, which will undoubtedly help us study more SNRs and candidates in the near future.

\begin{acknowledgements}
    Partly based on observations with the 100-m telescope of the MPIfR (Max-Planck-Institut f\"ur Radioastronomie) at Effelsberg.  HB acknowledges support from the European Re-search Council under the Horizon 2020 Framework Programme via the ERC Consolidator Grant CSF-648505. HB also acknowledges support from the Deutsche Forschungsgemeinschaft in the Collaborative Research Center (SFB 881) "The Milky Way System" (subproject B1).  NR acknowledges support from MPG through Max-Planck India partner group grant.  This research has made use of NASA's Astrophysics Data System and the SIMBAD database.  We have used the softwares Astropy \citep{2013A&A...558A..33A}, APLpy \citep{2012ascl.soft08017R},
    and DS9 \citep{2003ASPC..295..489J} at various stages of this research.  
\end{acknowledgements}

\bibliographystyle{aa.bst}          
\bibliography{ref}

\clearpage
\begin{appendix}

\onecolumn

\section{Images of the known SNRs studied in this work}
\label{apdx:knownSNRs}

\begin{figure*}[h]
    \centering
    \includegraphics[width=16cm]{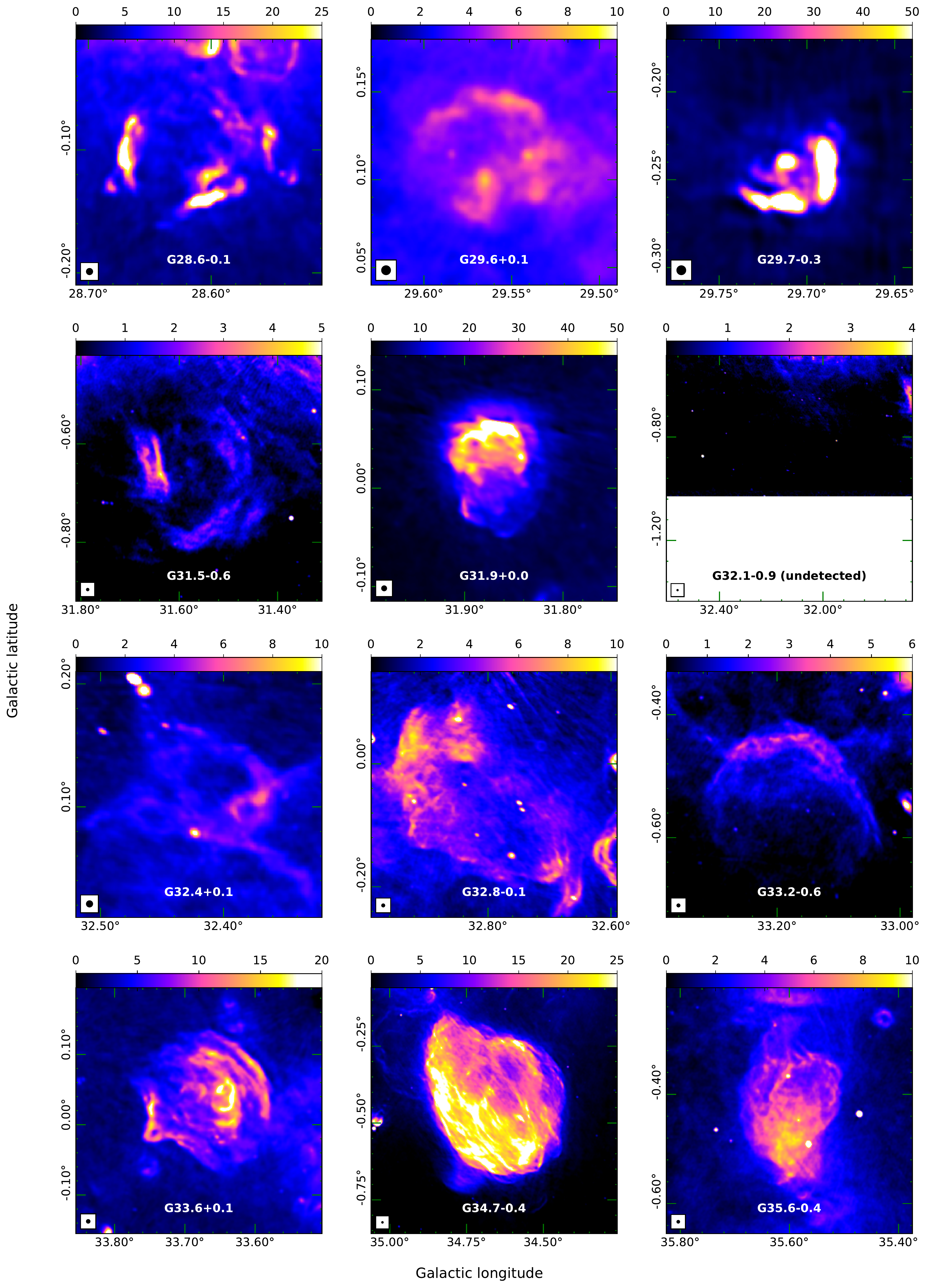}
    \caption{Total intensity maps of the GLOSTAR combination data of known SNRs in the pilot region in $\mathrm{mJy~beam}^{-1}$.}
    \label{figx:allknownI}
\end{figure*}

\begin{figure*}[h]
    \centering
    \includegraphics[width=16cm]{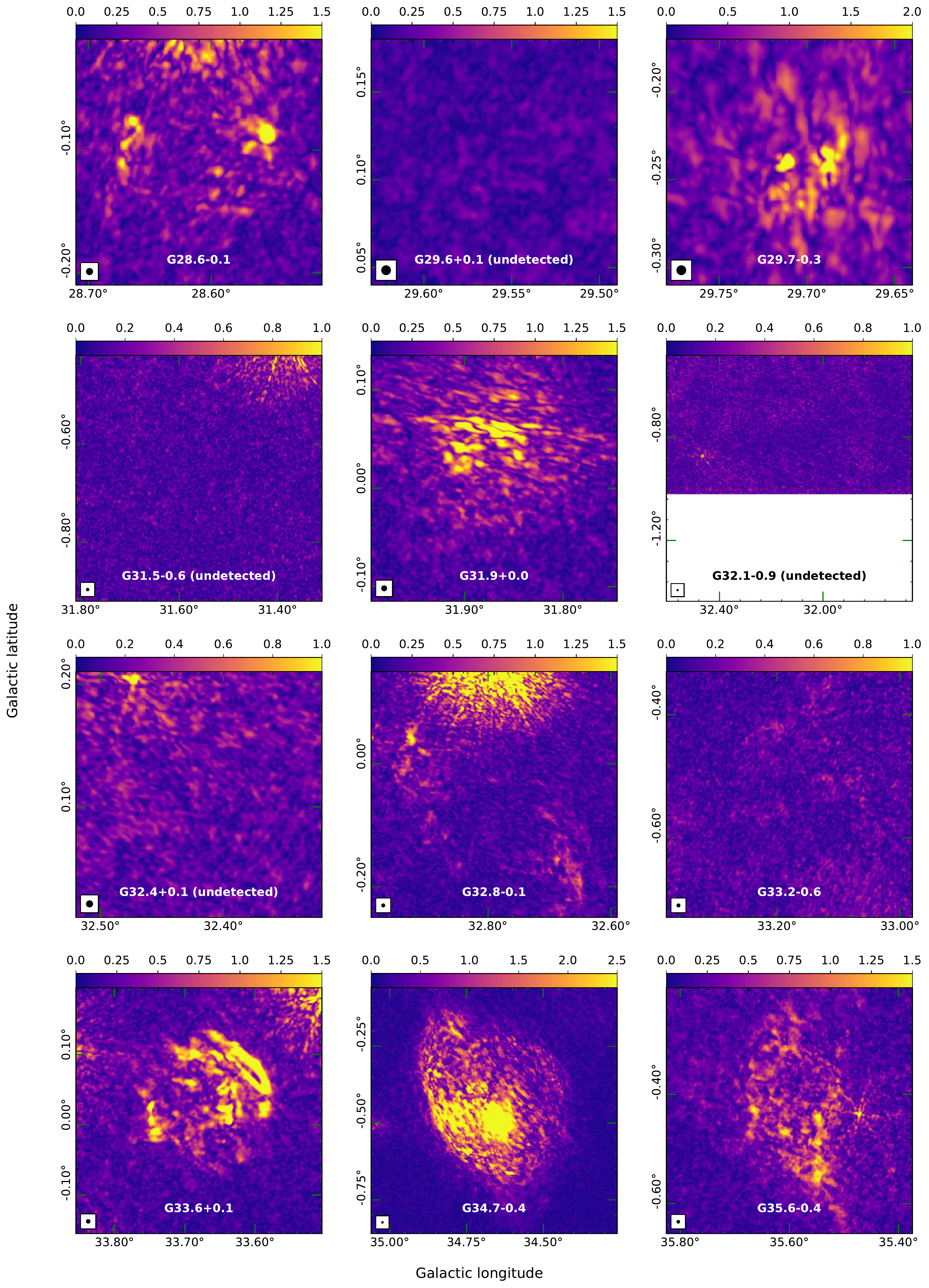}
    \caption{Linearly polarized intensity maps of the GLOSTAR combination data of known SNRs in the pilot region in $\mathrm{mJy~beam}^{-1}$.}
    \label{figx:allknownL}
\end{figure*}

\clearpage

\section{Images of the SNR candidates with an unclear morphology}

\begin{figure*}[h]
    \centering
    \includegraphics[width=11cm]{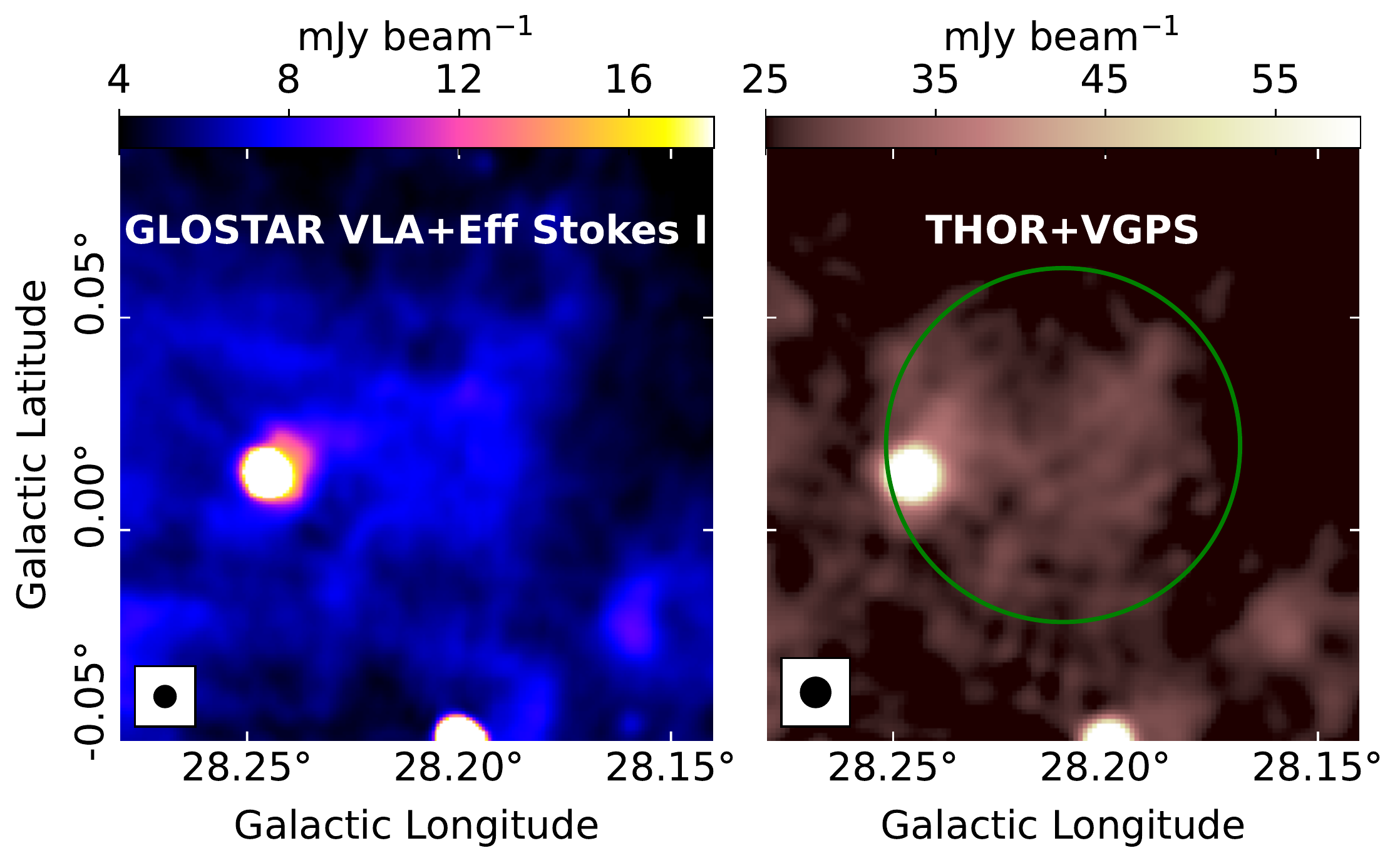}\\
    \includegraphics[width=11cm]{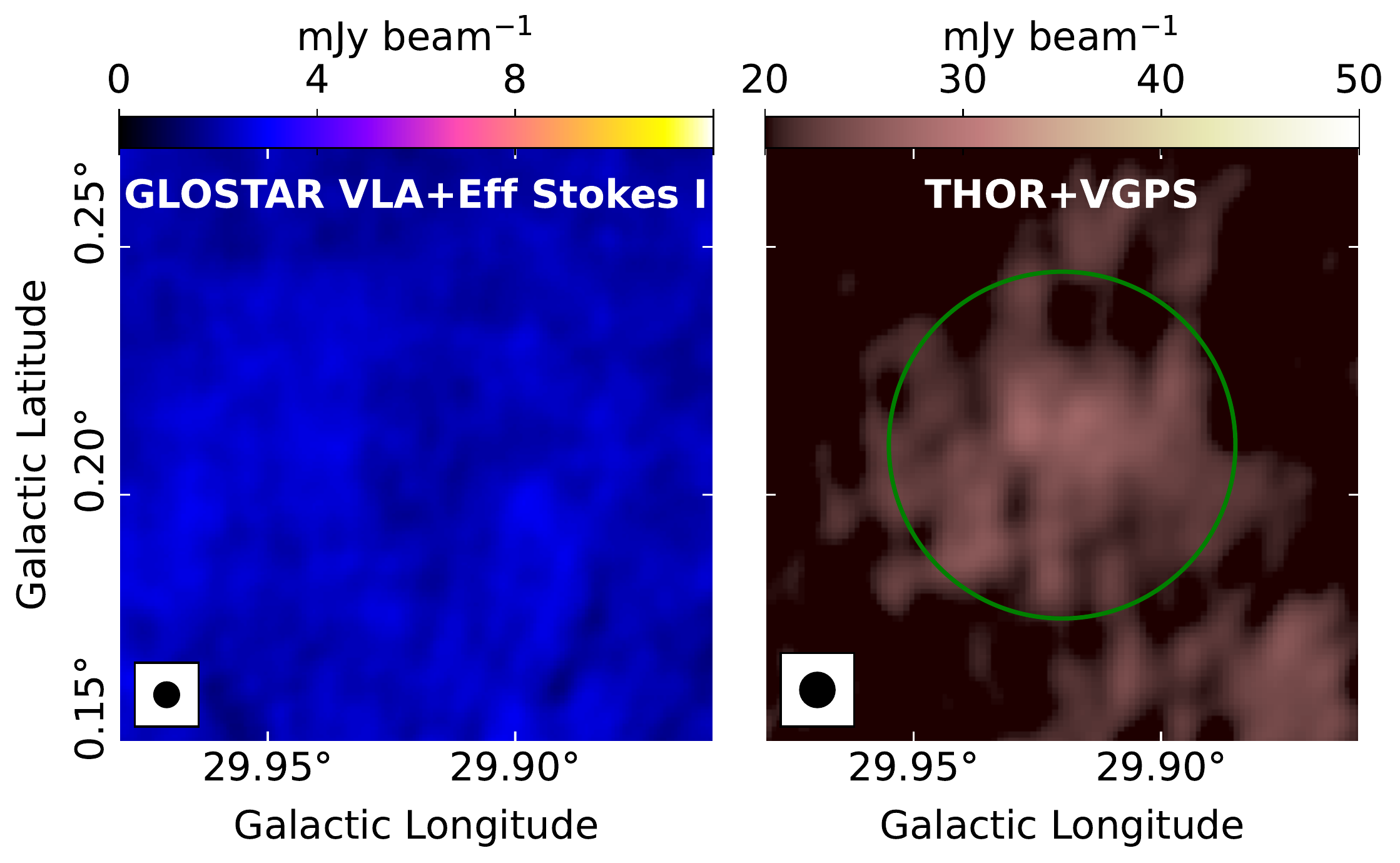}\\
    \includegraphics[width=11cm]{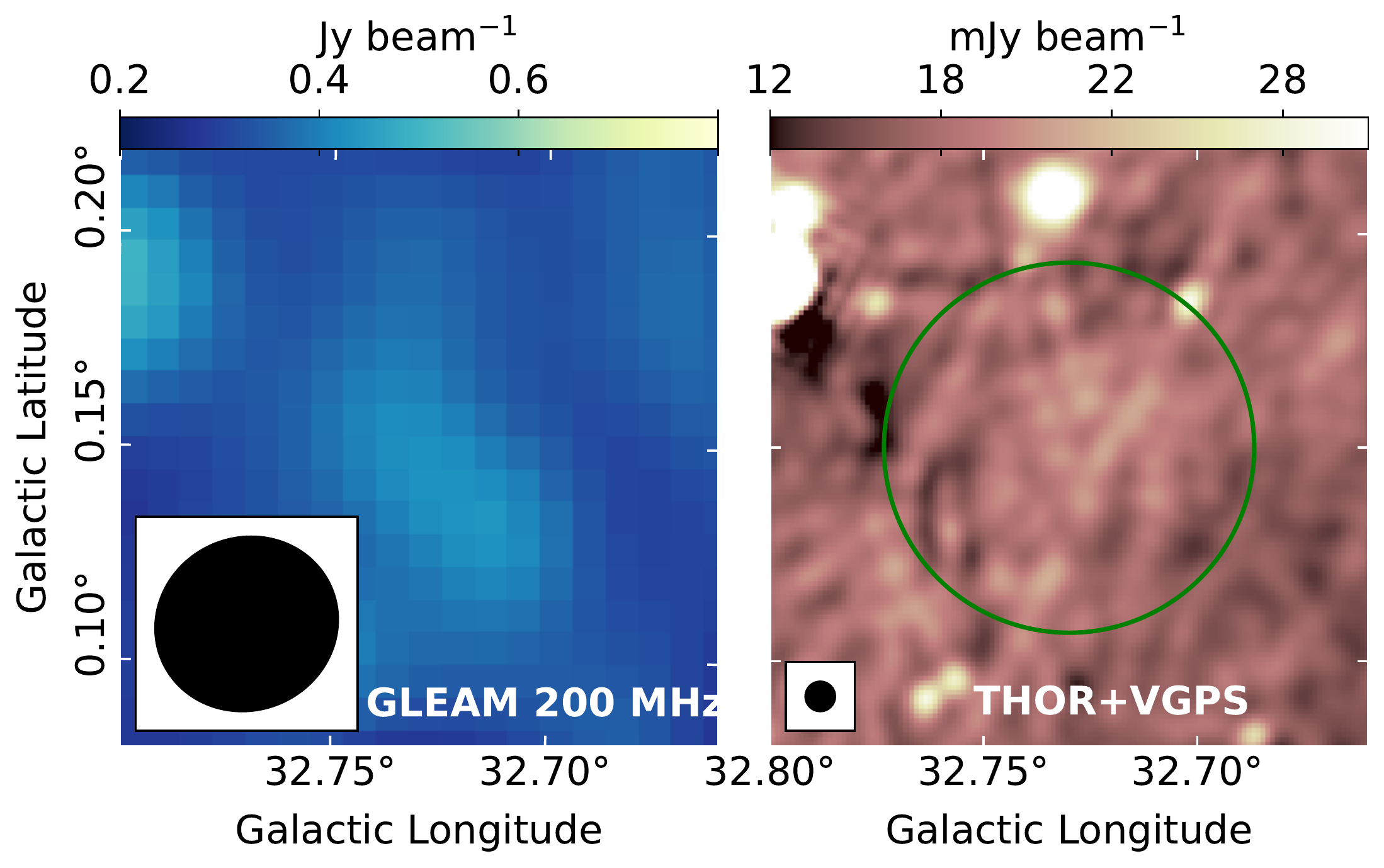}
    \caption{ Images of three SNR candidates from THOR: G28.21+0.02 (top panels), G29.92+0.21 (middle panels) and G32.73+0.15 (bottom panels).  The diffuse emission from G28.21+0.02 overlaps with the bright H~{\footnotesize{II}} region at $l{\sim}28.25\degree$, $b{\sim}0.01\degree$.  G29.92+0.21 and G32.73+0.15 are not detected in the GLOSTAR combination images.  }
    \label{fig:G28.21}
\end{figure*}

\begin{figure*}[h]
    \centering
    \includegraphics[width=\textwidth]{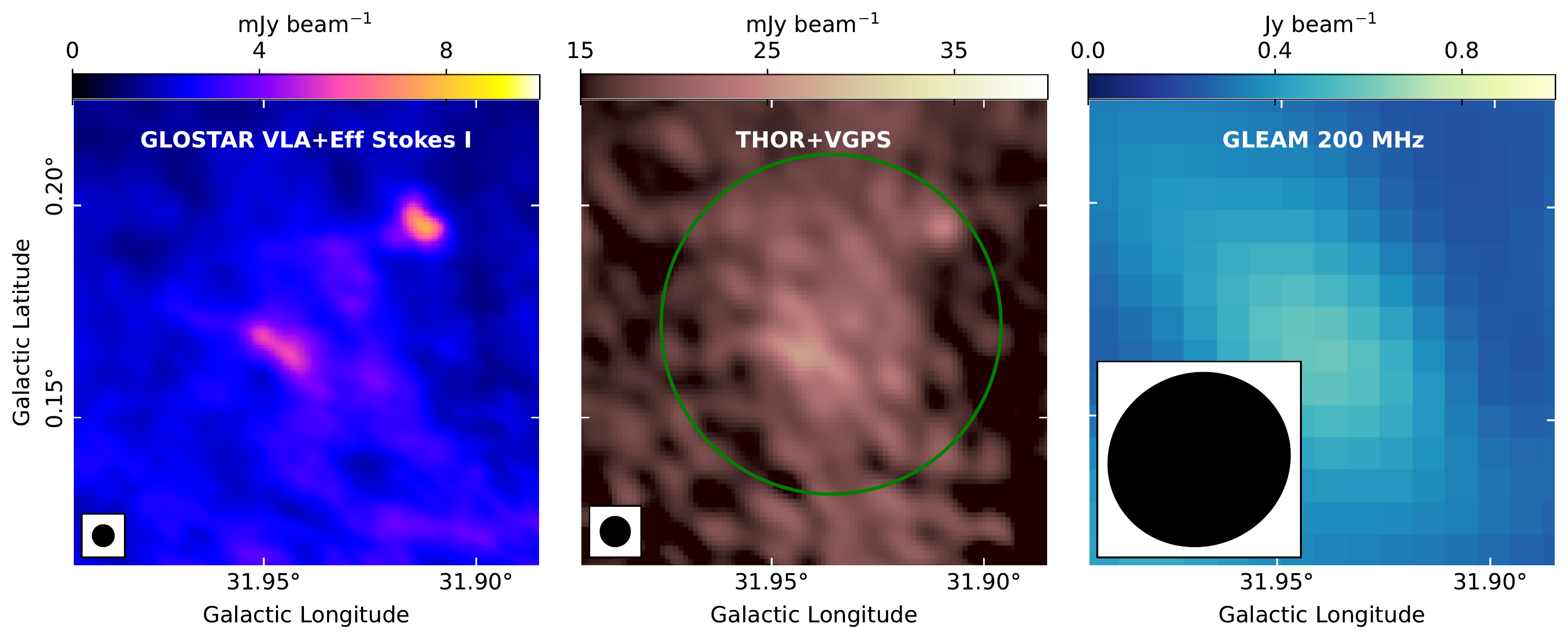}
    \caption{ G31.93+0.16, a SNR candidate from THOR, as seen in the GLOSTAR combination (left), the THOR+VGPS (middle) and the 200~MHz GLEAM data (right). }
    \label{fig:G31.93}
\end{figure*}

\end{appendix} 

\end{document}